\documentclass[12pt,prd,onecolumn,showpacs,amsmath,amssymb,aps,floats,floatfix,nofootinbib]{revtex4-1}

\usepackage[colorlinks=true,urlcolor=blue,anchorcolor=blue,citecolor=blue,filecolor=blue,linkcolor=blue,menucolor=blue,pagecolor=blue,linktocpage=true]{hyperref} 

\usepackage[inline]{enumitem}
\usepackage[multidot]{grffile} 
\usepackage{dcolumn}
\usepackage{bm}
\usepackage{amsmath}
\usepackage{amsfonts}
\usepackage{amssymb}
\usepackage{color}
\usepackage{latexsym}
\usepackage{slashed} 
\usepackage{pstricks}
\usepackage{indentfirst}
\usepackage{mathrsfs}
\usepackage{multirow}
\usepackage{epsfig,psfrag}
\usepackage{subfigure}
\usepackage{mathtools}
\usepackage{setspace} 
\usepackage[utf8]{inputenc} 
\usepackage[scientific-notation=true]{siunitx} 
\usepackage{verbatim}

\graphicspath{{fig/}}

\newcommand{\SUtwoL}{\mathrm{SU}(2)_\mathrm{L}}
\newcommand{\UoneY}{\mathrm{U}(1)_\mathrm{Y}}

\newcommand{\svann}{\left<\sigma_\mathrm{ann}v\right>}
\newcommand{\SUtwoD}{\mathrm{SU}(2)_\mathrm{D}}

\makeatother

\allowdisplaybreaks 

\begin{document}

\title{Hidden $\SUtwoD$ vector dark matter with a scalar septuplet}
\author{Zhao {\sc Zhang}}
\affiliation{School of Physics, Sun Yat-Sen University, Guangzhou 510275, China}
\author{Chengfeng {\sc Cai}}
\email{caichf3@mail.sysu.edu.cn}
\affiliation{School of Physics, Sun Yat-Sen University, Guangzhou 510275, China}
\author{Hong-Hao {\sc Zhang}}
\email{zhh98@mail.sysu.edu.cn}
\affiliation{School of Physics, Sun Yat-Sen University, Guangzhou 510275, China}

\begin{abstract}
We propose a vector dark matter model from a hidden $\SUtwoD$ gauge symmetry at $\si{TeV}$ scale.
A scalar septuplet is introduced to break the $\SUtwoD$ symmetry spontaneously. The septuplet also play the role of a portal between the standard model and the dark sector.
We find that there are two different vacuum configurations corresponding to the sign of the quartic coupling $\lambda_3$, which yields different mass spectrum for the gauge bosons. For a $\lambda_3<0$, the masses of gauge bosons are splitting, while for a $\lambda_3\geq0$, the masses are degenerate. We also study the RG evolutions of the couplings, and find that the perturbativity and vacuum stability can set a stringent bound on the parameter space.
For the phenomenological aspect, we consider the experimental constraints including dark matter direct detection, indirect detection, relic density, and Higgs couplings measurements. We find that there are parameter space survive from all the constraints, and they can be tested in future dark matter direct and indirect detection experiments. 
\end{abstract}

\maketitle
\tableofcontents

\section{Introduction}

Cosmology and astrophysics have provided many hints for the existence of dark matter (DM).
According to the observation of cosmic microwave background, the energy density of cold dark matter is about $\Omega_{\text{CDM}} h^2=0.1200\pm0.0012$~\cite{Planck:2018vyg}, which cannot be explained by the standard model (SM) of particle physics.
In order to unravel the DM mystery, people have tried many scenarios beyond SM.
One of the most famous paradigms is the weakly interacting massive particles (WIMPs), which can realize the observed DM relic density through the thermal freeze-out mechanism.
Based on the interaction properties of WIMPs, people have designed many underground experiments aiming to directly detect the recoil signal from the collision of WIMP and nucleus in detectors~\cite{LUX:2016ggv,XENON:2018voc,PandaX-4T:2021bab,LUX-ZEPLIN:2022qhg} . However, no convincing signal has been detected so far, resulting in a stringent bound on the DM-nucleon cross section.

From the view of particle physics, the WIMP DM candidate can be an elementary particle with various spins.
Tons of scalar and fermionic DM models have been studied in literatures, while the vector DM models are less explored.
The most natural way to introduce an elementary vector DM field is basing on a new gauge symmetry, which can be abelian or non-abelian.
For a gauged U(1) vector DM model, a discrete symmetry should be imposed manually in order to stabilize the DM particle~\cite{Lebedev:2011iq,Abe:2012hb,Farzan:2012hh,Baek:2012se,Domingo:2013tna,Yu:2014pra,Chen:2014cbt,Duch:2015jta,DiFranzo:2015nli,Azevedo:2018oxv,Mohamadnejad:2019vzg,Glaus:2019itb,Arcadi:2020jqf,Delaunay:2020vdb,Salehian:2020asa}.
On the other hand, in the non-abelian group cases, the discrete symmetry can be induced as a remnant of spontaneous symmetry breaking (SSB) naturally, and thus the vector bosons can be stable.
The simplest setup of non-abelian model is to consider a gauged $\SUtwoD$ with some multiplet fields, such as a scalar doublet~\cite{Hambye:2008bq,Boehm:2014bia,Gross:2015cwa,Karam:2015jta,Khoze:2016zfi,Baouche:2021wwa,Borah:2021ftr}, a scalar triplet~\cite{Baek:2013dwa,Khoze:2014woa,Ghosh:2020ipy,Nomura:2020zlm}, two scalar triplets~\cite{Ghosh:2020ipy,Hu:2021pln}, scalar doublet and triplet~\cite{Cai:2021wmu}, scalar and Dirac fermion doublet~\cite{Belyaev:2022shr,Elahi:2022hgj}, scalar triplet and quintet~\cite{Nomura:2020zlm}, etc~\cite{Barman:2018esi,Barman:2019lvm,Chowdhury:2021tnm}. The multiplets can be a portal to the SM through a renormalizable or non-renormalizable operators.

A real inert scalar septuplet under the SM $\SUtwoL$ in the framework of minimal dark matter model~\cite{Cirelli:2005uq} had been studied in Ref.~\cite{Cai:2015kpa}. In that case, the septuplet does not obtain any vacuum expectation value (VEV), and the neutral component of the septuplet is regarded as the DM candidate. It has been proved that the quartic couplings of the septuplet in this model have Landau pole below the 
 scale due to the high dimensional representation~\cite{Hamada:2015bra,Cai:2015kpa}. 
In this work, we consider a dark sector to be a new $\SUtwoD$ gauge theory with a real scalar septuplet. The septuplet can obtain a nonzero VEV for breaking the $\SUtwoD$ spontaneously and making the gauge bosons massive.
We will prove that there are two different vacuum configurations, corresponding to different sign of the quartic coupling $\lambda_3$. In addition, after the SSB of $\SUtwoD$, there are remnant discrete symmetries, $Z_6 \times Z_2$ and $Z_4 \times Z_2$ in these two cases respectively, which keep the vector DM stable. 
The vector DMs interact with SM particles through Higgs-portal and thus the vector DMs can be the cold dark matter candidate which produced via the freeze-out mechanism. The direct detection~\cite{LUX-ZEPLIN:2022qhg,LZ:2015kxe} and indirect detection~\cite{MAGIC:2016xys,CTAConsortium:2012fwj} bounds for WIMPs DM can be used to constrain the model as well. In addition, with the help of the threshold effect~\cite{Elias-Miro:2012eoi}, the portal interaction can help to alleviate the problem of unstable vacuum generated by the SM Higgs quartic coupling.
Since the gauge coupling of $\SUtwoD$ can be smaller than the one of $\SUtwoL$, the Landau pole problem can be avoided in our model. On the other hand, the requirement of coupling perturbativity and vacuum stability is capable to put stringent theoretical constraints on the the parameter space.

The paper is organized as follows.
In Sec.~\ref{sec:model}, we will introduce the setup of our model, including the vacuum configurations and the mass spectra of particles.
The RG running of the couplings and the vacuum stability conditions are introduced in Sec.~\ref{sec:VSLP}.
The constraints from DM phenomenologies, Higgs couplings measurements, perturbativity and vacuum stability will be presented in Sec.~\ref{sec:DM}.
Finally, we will conclude and discuss our result in Sec.~\ref{sec:conclusion} and give the analytic solution of Landau pole in Appendix~\ref{sec:append}. 

\section{The setup of the model}
\label{sec:model}
In Ref.~\cite{Cai:2021wmu}, a vector dark matter has been studied under the framework of gauged $\SUtwoD$ completely broken by two real triplets. In that case, a splitting mass spectrum of the vector fields can be obtained, however there are many couplings in the potential and make the model less predictive. In this work, we consider the gauged $\SUtwoD$ symmetry to be completely broken by a scalar septuplet. This model has a much simpler potential term, but the vacuum structures are non-trivial. We will show that there are two different of vacuum configurations depending on the sign of a quartic coupling. The gauge bosons, which are assumed to play the role of cold dark matter candidates, have different mass spectrum in these different cases.

\subsection{The 7-dimensional representation of $\SUtwoD$ and the Lagrangian}
The 7-dimensional representation of $\mathrm{SU}(2)$ generators are defined by
\begin{eqnarray}
\tau^1=\frac{\tau^++\tau^-}{2},\quad \tau^2=\frac{\tau^+-\tau^-}{2i},\quad \tau^3=\text{diag}\{3,2,1,0,-1,-2,-3\},
\end{eqnarray}
where
\begin{eqnarray}
\tau^+=\begin{pmatrix}
0&-\sqrt{6}&&&&&\\
&0&-\sqrt{10}&&&&\\
&&0&-\sqrt{12}&&&\\
&&&0&\sqrt{12}&&\\
&&&&0&\sqrt{10}&\\
&&&&&0&\sqrt{6}\\
&&&&&&0\\
\end{pmatrix},\quad \tau^-=(\tau^+)^\text{T},
\end{eqnarray}

It is well known that there exists a unitary symmetric (anti-symmetric) matrix $S$ transforming the $\mathrm{SU}(2)$ generators into their complex conjugation as follows 
\begin{eqnarray}
	S_{(n)} \tau^a S^{-1}_{(n)} = -(\tau^a)^*,\quad \mathrm{where}\quad
	S_{(n)ij}=\left\{ \begin{array}{rl}
	\delta_{i, n + 1 - j} & \qquad\text{if}\quad i \leqslant \frac{n}{2},\\
	(- 1)^{n + 1} \delta_{i, n + 1 - j} & \qquad\text{if}\quad i > \frac{n}{2}.
	\end{array} \right.
\end{eqnarray}
The $\Phi$ field transforms as $\Phi\to S\Phi$ correspondingly, behaves like a field in the complex conjugate representation. If $\Phi^\ast=S\Phi$, then $\Phi$ is said to be a real scalar.
A real septuplet $\Phi$ can be written as
\begin{equation}
\Phi = \frac{1}{\sqrt{2}}  (\Delta_3, \Delta_2, \Delta_1, \Delta_0,
\Delta_{- 1}, \Delta_{- 2}, \Delta_{- 3})^\text{T},
\end{equation}
where $\Delta_i^*=\Delta_{-i}$.

The potential terms for the scalar fields are given by
\begin{eqnarray}
V (H, \Phi) & = & - \mu^2 H^{\dag} H + \lambda (H^{\dag} H)^2 - m^2
\Phi^{\dag} \Phi + \lambda_2 (\Phi^{\dag} \Phi)^2 + \frac{\lambda_3}{48}
(\Phi^{\dag} \tau^a \tau^b \Phi)^2 + \lambda_4 (H^{\dag} H) (\Phi^{\dag} \Phi)\notag \\
& = & - \mu^2 H^{\dag} H + \lambda (H^{\dag} H)^2 - m^2 \Phi^{\dag} \Phi +
\lambda'_2 (\Phi^{\dag} \Phi)^2 + \frac{\lambda_3}{48} (\Phi^{\dag} T^{a b}
\Phi)^2 + \lambda_4 (H^{\dag} H) (\Phi^{\dag} \Phi). \label{eq:V}\nonumber\\
\end{eqnarray}
where $H$ is the SM $\SUtwoL$ Higgs doublet.
In the second line, we have separated out the traceless part of $\tau^a \tau^b$ and define $T^{ab} \equiv \{\tau^a, \tau^b\}/2 - 4 \delta^{ab}$, $\lambda_2' \equiv \lambda_2 + \lambda_3$. The explicit expression of the $\lambda_3$ term in Eq.\eqref{eq:V} is given as follows, 
\begin{eqnarray}\label{lam3term}
& &\quad\frac{1}{48}  (\Phi^{\dag} T^{a b} \Phi)^2 \nonumber\\
& =&  \frac{1}{8} \Delta_0^4 +
\frac{21}{32}  (\Delta_{- 1} \Delta_1)^2 + \frac{25}{32}  (\Delta_{- 3}
\Delta_3)^2 + \frac{1}{2} \Delta_0^2 \Delta_{- 1} \Delta_1 + \frac{5}{4}
\Delta_0^2 \Delta_{- 2} \Delta_2 - \frac{5}{8} \Delta_0^2 \Delta_{- 3}
\Delta_3 \nonumber\\
&\quad& +\frac{15}{16} \Delta_{- 1} \Delta_1 \Delta_{- 2} \Delta_2 -
\frac{5}{16} \Delta_{- 1} \Delta_1 \Delta_{- 3} \Delta_3 + \frac{25}{16}
\Delta_{- 2} \Delta_2 \Delta_{- 3} \Delta_3+ \nonumber\\
&\quad & \left( - \frac{\sqrt{15}}{8} \Delta_1^3 \Delta_{- 3} -
\frac{\sqrt{30}}{16} \Delta_0 \Delta_1^2 \Delta_{- 2} + \frac{5
	\sqrt{15}}{16} \Delta_1 \Delta_{- 2}^2 \Delta_3 + \frac{15 \sqrt{2}}{16}
\Delta_0 \Delta_1 \Delta_2 \Delta_{- 3} + \text{h.c.} \right) . 
\end{eqnarray}

The $\lambda_4$ term in Eq.\eqref{eq:V} is a portal connecting the SM and the dark sectors, and thus all the DM phenomenologies relate to it. It is also responsible for prevent the vacuum becomes unstable in the high energy by raising the self-coupling of the SM Higgs via the threshold effect.

The kinetic terms of our dark sector are given by
\begin{align}
\mathcal{L}_{\text{kin}} &= -\frac{1}{4} F^a_{\mu\nu}F^{a \mu\nu} + (D^{\mu} \Phi)^{\dag} D_{\mu} \Phi \notag\\
&=\frac{1}{2} X^{a \mu} (g_{\mu \nu} \partial^2 -
\partial_{\mu} \partial_{\nu}) X^{a \nu} - g_4 \varepsilon^{a b c}  (\partial_{\mu} X_{\nu}^a) X^{b\mu} X^{c \nu} - \frac{1}{4} g_4^2 \varepsilon^{a b c} \varepsilon^{a d e} X^b_{\mu} X^c_{\nu} X^{d \mu} X^{e \nu} \notag\\
&\quad + (\partial^{\mu} \Phi^{\dag}) \partial_{\mu} \Phi + i g_4 X^a_{\mu}
[\Phi^{\dag} \tau^a \partial_{\mu} \Phi - (\partial_{\mu} \Phi^{\dag})
\tau^a \Phi] + g_4^2 X^a_{\mu} X^{b \mu} \Phi^{\dag} \tau^a \tau^b \Phi
,\label{eq:kinetic}
\end{align}
where $F^a_{\mu\nu}$ is the field strength tensor of gauge bosons $X^a$, $D_\mu=\partial_\mu - i g_4 X^a_\mu \tau^a$ is the covariant derivative and $g_4$ is the gauge coupling coefficient.

In this work, we consider the lightest gauge bosons $X^a_\mu$ as WIMP DM candidate.
They become massive after the septuplet breaking $\SUtwoD$ symmetry completely, which are presented in next subsection.
The VEV of septuplet $v^\prime$ is set around a few $\si{TeV}$ to $10~\si{TeV}$. Some discrete symmetries is retained and make the DM stable. 

\subsection{Vacuum configurations and mass spectrum}

The vacuum configurations of the septuplet can be obtained by minimizing the the potential term in Eq.\eqref{eq:V}. We only consider the tree level potential for simplicity, and assume that it could drive the spontaneous symmetry breaking of SU(2)$_L\times$U(1)$_Y$ and $\SUtwoD$.  Before proceeding, we want to make some comments on the $\lambda_3$ term in the potential, Eq.\eqref{eq:V}. If this term is absent, the tree-level potential respects a global SO(7) symmetry. A negative mass-squared term for $\Phi$ induce the spontaneous breaking from SO(7) to SO(6), and then there is a 6-dimensional degenerate vacuum (the coset SO(7)/SO(6) is an $S^6$) corresponding to six Goldstones. Three (or two) of the Goldstones become the longitudinal polarizations of the gauge fields, while there still remains three (or four) flat directions for the potential. Radiative corrections from gauge boson loops can lift the potential since gauge interactions explicitly break the SO(7), and then the vacuum is aligned in the $\Delta_0$ direction. It implies that the gauged $\SUtwoD$ can not be completely broken in this case, and a U(1) subgroup is restored. If the $\lambda_3$ term in Eq.\eqref{eq:V} is turned on, the SO(7) is explicitly broken into an SO(3) ( SU(2) ) subgroup. It leads to a vacuum configuration which completely breaks $\SUtwoD$.

In a generic case, every components of the septuplet can acquire nonzero VEV
\begin{eqnarray}
\Phi = \frac{1}{\sqrt{2}}  \left( v_3, v_2, v_1, v_0, v_1^*,
v_2^*, v_3^* \right)^\text{T}. \label{eq:gVEV}
\end{eqnarray}
If we ignore the $\lambda_3$ terms in the potential, the potential is a function of radial component $v^\prime$ defined by $v^\prime \equiv \sqrt{v_0^2 + 2| v_1 |^2	+ 2| v_2 |^2 +2 | v_3 |^2}$. We can perform a $\text{SO} (7)$ rotation
to keep only one of 7 components nonzero.
The potential is flat along a $S^6$ perpendicular to the radial direction, which means it only depends on $v^\prime$.
Since the $\lambda_3$ term explicitly breaks the SO(7) symmetry to SU(2), the potential becomes a function of the $S^6$ coordinates. Therefore, for a fixed $v^\prime$, the $\lambda_3$ term forces the minima to locate in some specific directions of to the $S^6$.

We can use the $\SUtwoD$ transformation to eliminate some degrees of freedom in Eq.~\eqref{eq:gVEV}.
Applying a transformation $U = e^{i \frac{\pi}{2} \tau_1} e^{i \varphi \tau^3}$ to $\Phi$ as $\Phi' = U\Phi$, the $v_0$ component becomes
\begin{eqnarray}
v_0' = - \frac{\sqrt{5}}{2} | v_3 | \left[ \sin (3 \varphi +
\varphi_3) + \sqrt{\frac{3}{5}} \frac{| v_1 |}{| v_3 |} \sin (\varphi +
\varphi_1) \right].
\end{eqnarray}
where the $\varphi_1$ and $\varphi_3$ are the complex phase angles of $v_1$ and $v_3$, respectively.
We can tune $\varphi$ to set the expression in square bracket vanishing and eliminate the $v_0$ component.
We can further perform another $\SUtwoD$ transformation $e^{i \varphi'\tau_3}$ after $U$ to eliminate the complex phase of the $v_1$ component, and then the vacuum can be rewritten as
\begin{eqnarray}\label{vev_conf}
\Phi = \frac{1}{\sqrt{2}}  (v_3, v_2, v_1, 0, v_1, v_2^*, v_3^*)^\text{T} .
\end{eqnarray}

Substituting Eq.\eqref{vev_conf} into $ (\Phi^{\dag} T^{a b} \Phi)^2$, one obtains
\begin{eqnarray}
(\Phi^{\dag} T^{a b} \Phi)^2 & = & \frac{1}{4} \left[ 3 v_1^2 + 5 |v_3|^2 - \sqrt{15} v_1 (v_3 + v_3^*)\right]^2  \notag\\
&  &+ \frac{1}{4} \left[ 9 v_1^2 - 5 |v_3|^2 - \sqrt{15} v_1
(v_3 + v_3^*) \right]^2 \notag \\
&  & + (3 v_1^2 - 5 |v_3|^2)^2 - \frac{15}{2} v_1^2 (v_3 -
v_3^*)^2 \notag\\
&  & + \frac{1}{8} \left[ 3 \sqrt{10} v_1 (v_2 + v_2^*) + 5 \sqrt{6}
(v_2^* v_3 + v_2 v_3^*) \right]^2  \notag\\
&  & - \frac{1}{8} \left[ 3 \sqrt{10} v_1 (v_2 - v_2^*) + 5 \sqrt{6}
(v_2^* v_3 - v_2 v_3^*) \right]^2  \notag\\
& = & \frac{63}{2} v_1^4 + \frac{75}{2} |v_3|^4 - 15 v_1^2 |v_3|^2 - 6 \sqrt{15} v_1^3 (v_3 + v_3^*)  \notag\\
&  & + 45 v_1^2 |v_2|^2 + 75 |v_2|^2 |v_3|^2 + 15 \sqrt{15} v_1 (v_2^2 v_3^* + v_2^{* 2} v_3) . 
\end{eqnarray}
Note that $(\Phi^{\dag} T^{a b} \Phi)^2$ is a sum of squares and thus is non-negative.
The first three lines after the first equal sign depend on the $v_1$ and $v_3$, while the last two lines after the first equal sign depend on $v_1$, $v_2$ and $v_3$.
We find that there are two inequivalent vacuum configurations corresponding to different signs of $\lambda_3$:
\begin{itemize}
\item For $\lambda_3 \ge 0$, it is easy to see that solutions $(|v_2| = v'/\sqrt{2}, v_1 = v_3 = 0)$ and $( v_2 = 0, v_3 = \sqrt{15}v_1/5 )$ minimize the potential since $(\Phi^{\dag} T^{a b} \Phi)^2$ becomes zero.
These two solutions are equivalent and related by a specific $\SUtwoD$ transformation $e^{- i (\pi/2) \tau_2} e^{- i (\text{Arg}(v_2)/2) \tau_3}$. Actually, there are infinite equivalent solutions generated by applying a continuous $\SUtwoD$ transformation on $(|v_2| = v'/\sqrt{2}, v_1 = v_3 = 0)$.
\item For $\lambda_3 < 0$, to find the minimum of the potential, we should maximize $(\Phi^{\dag} T^{a b} \Phi)^2$ with a fixing $v' = \sqrt{2 v_1^2 + 2|v_2|^2 + 2|v_3|^2}$.
Let us rewrite $(\Phi^{\dag} T^{a b} \Phi)^2$ with $v_2 = | v_2 | e^{i \varphi_2}$ and $v_3 = | v_3 | e^{i \varphi_3}$,
\begin{eqnarray}
	(\Phi^{\dag} T^{a b} \Phi)^2 & = & \frac{63}{2} v_1^4 + \frac{75}{2} |v_3|^4 - 15 v_1^2 | v_3 |^2 - 12 \sqrt{15} v_1^3 | v_3 | \cos
	(\varphi_3)\nonumber\\
	&  & + 45 v_1^2 | v_2 |^2 + 75 |v_2|^2 |v_3|^2 + 30 \sqrt{15} v_1 | v_2 |^2 | v_3 | \cos (2 \varphi_2 -
	\varphi_3) .
\end{eqnarray}
Obviously, this term is maximized when $\varphi_3 = \pi$ and $\varphi_2 = \pi/2$, so we can pin down these two angles.
Next, we can rewrite $| v_2 | = v'\cos \theta/\sqrt{2}$, $v_1 = v'\sin \theta \sin \phi/\sqrt{2}$, $|v_3| = v'\sin \theta \cos \phi/\sqrt{2}$, yielding
\begin{eqnarray}
	(\Phi^{\dag} T^{a b} \Phi)^2 &=& {v'}^4 \left( \frac{63}{8} \sin^4
	\theta \sin^4 \phi + \frac{75}{8} \sin^4 \theta \cos^4 \phi - \frac{15}{4}
	\sin^4 \theta \sin^2 \phi \cos^2 \phi \right. \nonumber\\
	&&\quad + 3 \sqrt{15} \sin^4 \theta \sin^3
	\phi \cos \phi + \frac{45}{4} \cos^2 \theta \sin^2 \theta \sin^2 \phi +
	\frac{75}{4} \cos^2 \theta \sin^2 \theta \cos^2 \phi \nonumber\\
	&&\quad+\left. \frac{15\sqrt{15}}{2} \cos^2 \theta \sin^2 \theta \sin \phi \cos \phi \right).
\end{eqnarray}
This function reaches its maximum $75/8$  at a specific configuration $(v_1 = v_2 = 0, v_3= v'/\sqrt{2})$. The other equivalent vacuum, e.g. $\left( \theta = \pi/2, \tan \phi = \sqrt{15} \right)$ can be found by applying an $\SUtwoD$ transformation on the specific vacuum. 
\end{itemize}
After spontaneous symmetry breaking, the gauge bosons become massive.
We can now study the mass spectrum for these two inequivalent vacua.

\subsubsection{The $\lambda_3<0$ case (splitting spectrum)}
For the $\lambda_3<0$ case, we consider the vacuum configuration to be $\Phi = \left( v'/2, 0, 0, 0, 0, 0, v'/2 \right)^\text{T}$.
The mass-squared of gauge bosons are
\begin{eqnarray}
m_{X^{\pm}}^2 = \frac{3 g_4^2}{2} {v'}^2, \quad m_{Z'}^2 = 9 g_4^2 {v'}^2,
\end{eqnarray}
where $X^\pm \equiv (X^1 \mp i X^2)/\sqrt{2}$, $Z'\equiv X^3$. In this case, there are mass splitting between $X^\pm$ and $Z^\prime$. Since $Z^\prime$ is heavier than twice of $X^\pm$'s mass, it can decay into a pair of $X^\pm$.

The Goldstone bosons eaten by gauge bosons are $\Delta_2$ and $-\textrm{Im}(\Delta_3)$. 
The physical scalar fields are $\Delta_0$, $\Delta_1$ and a mixture of the neutral $CP$-even component, $h$ of SM Higgs doublets, and the CP-even mode around the VEV of $\Delta_3$.
We can find their mass-squared as follows
\begin{eqnarray}
m_{\Delta_0}^2 = - \frac{45\lambda_3}{32} {v'}^2, \quad m_{\Delta_1}^2 = - \frac{15 \lambda_3}{16} {v'}^2,
\end{eqnarray}
and
\begin{eqnarray}
\mathcal{L}_{M_{\Delta_3}} & = & - \frac{1}{2}  \left( \begin{array}{cc} h & \rho_3
\end{array} \right) 
\left( \begin{array}{cc} 2 \lambda v^2 & \lambda_4 v v'\\
\lambda_4 v v' & \left( 2 \lambda'_2 + \frac{25 \lambda_3}{16} \right){v'}^2
\end{array} \right) 
\left( \begin{array}{c} h\\\rho_3\end{array} \right) \notag\\
& = & - \frac{1}{2} \left[ \lambda v^2 + \left( \lambda_2 + \frac{25
	\lambda_3}{32} \right) {v'}^2 - \sqrt{\left[ \lambda v^2 - \left( \lambda'_2
	+ \frac{25 \lambda_3}{32} \right) {v'}^2 \right]^2 + \lambda_4^2 v^2 v'^2} \right] h_1^2 \notag\\
&  & - \frac{1}{2} \left[ \lambda v^2 + \left( \lambda_2 + \frac{25
	\lambda_3}{32} \right) {v'}^2 + \sqrt{\left[ \lambda v^2 - \left( \lambda'_2
	+ \frac{25 \lambda_3}{32} \right) {v'}^2 \right]^2 + \lambda_4^2 v^2 v'^2} \right] h_2^2,
\end{eqnarray}
where $v\approx246~\si{GeV}$ is the electroweak VEV of SM Higgs doublet and the mass eigenstates $(h_1,~h_2)$ relates to the definition states by a rotation
\begin{eqnarray}
\left( \begin{array}{c}h\\\rho_3\end{array} \right) =
\left( \begin{array}{cc}\cos \alpha & \sin \alpha\\- \sin \alpha & \cos \alpha
\end{array} \right) \left( \begin{array}{c}h_1\\h_2\end{array} \right),
\end{eqnarray}
where
\begin{eqnarray}
\tan \alpha = \frac{\sqrt{\left[ \left( \lambda'_2 + \frac{25}{32}
			\lambda_3 \right) {v'}^2 - \lambda v^2 \right]^2 + (\lambda_4 v v')^2} -
		\left[ \left( \lambda'_2 + \frac{25}{32} \lambda_3 \right) {v'}^2 - \lambda
		v^2 \right]}{\lambda_4 v v'}.\label{eq:alpha_v3}
\end{eqnarray}
If we identify $h_1$ as the $125~\si{GeV}$ SM-like Higgs field, then $\lambda$ and the mass square of $h_2$ can be expressed as
\begin{eqnarray}
\lambda & \approx & \frac{m_{h_1}^2}{2 v^2} + \frac{\lambda_4^2 {v'}^2}{2 \left[ 2
	\left( \lambda'_2 + \frac{25}{32} \lambda_3 \right) {v'}^2 - m_{h_1}^2
	\right]}, \\
m_{h_2}^2 & \approx & 2 \left( \lambda'_2 + \frac{25}{32} \lambda_3 \right) {v'}^2
+ \frac{\lambda_4^2 v^2 {v'}^2}{2 \left( \lambda'_2 + \frac{25}{32}
	\lambda_3 \right) {v'}^2 - m_{h_1}^2}.\label{eq:mh2_v2} 
\end{eqnarray}

In this work, we only focus on the situation that the scalar $\Delta_0$ and $\Delta_1$ are massive enough for decaying into the gauge bosons. It requires the parameters to satisfy the condition that 
\begin{eqnarray}
\lambda_3 \le -48 g_4^2 /5~,\label{VDMcond1}
\end{eqnarray}
and thus complex vector boson $X^\pm$ is the lightest stable particle in the dark sector. 
Since we will regard the $X^\pm$ (and $X^\ast$) as the dark matter candidate in this case, we should ensure that they stable. We find that there is a remnant $Z_6 \times Z_2$ symmetry in the dark sector after the $\SUtwoD$ symmetry breaking:
\begin{eqnarray}
Z_6 & : & \Delta_0 \rightarrow -\Delta_0,\quad \Delta_{1} \rightarrow e^{-i \frac{2 \pi}{3}} \Delta_{1}, \quad \Delta_2 \rightarrow e^{i \frac{5 \pi}{3}} \Delta_2, \quad \Delta_3 \rightarrow \Delta_3, \notag\\
&  &  X^+ \rightarrow e^{-i \frac{5 \pi}{3}} X^+, \quad Z' \rightarrow Z' . \\
Z_2 & : & \Delta_0 \rightarrow -\Delta_0 \quad \Delta_1 \rightarrow  \Delta_{- 1},
\quad \Delta_2 \rightarrow -\Delta_{- 2}, \quad \Delta_3 \rightarrow \Delta_{- 3}, \notag\\
&  & X^+ \rightarrow X^-, \quad Z' \rightarrow - Z'
\end{eqnarray}
which can prevent the dark matter candidate $X^\pm$ from decaying to SM particles.

\subsubsection{The $\lambda_3\ge0$ case (degenerate spectrum)}
For the $\lambda_3\ge0$ case, we define $\Phi = \left( 0, v'/2, 0, 0, 0, v'/2, 0 \right)^\text{T}$.
The mass-squared of gauge bosons are given by
\begin{eqnarray}
m_{X^\pm}^2 = m_{Z'}^2 = 4 g_4^2 {v'}^2.
\end{eqnarray}
We can see that the masses of $X^\pm$ and $Z^\prime$ are degenerate in this case.

One of the Goldstone bosons is $-\textrm{Im}(\Delta_2)$, the other one is a mixture of $\Delta_1$ and $\Delta_3$
\begin{eqnarray}
\mathcal{L}_{M_{\Delta_{1, 3}}} & = & - \frac{5\lambda_3 {v'}^2}{32}  \begin{pmatrix}
\Delta_{- 1} & \Delta_3
\end{pmatrix} \begin{pmatrix} 3 &  \sqrt{15}\\
 \sqrt{15} & 5\end{pmatrix}  \begin{pmatrix}\Delta_1\\\Delta_{- 3}\end{pmatrix}  \nonumber\\
& = & - \begin{pmatrix}
G^+ & \Delta^{\ast}_{13}
\end{pmatrix}  \begin{pmatrix}
0 & \\
& \frac{5 \lambda_3 {v'}^2}{4}
\end{pmatrix} \begin{pmatrix}
G^-\\
\Delta_{13}
\end{pmatrix} , 
\end{eqnarray}
with
\begin{eqnarray}
\begin{pmatrix}\Delta_1\\\Delta_{- 3}\end{pmatrix} 
= \begin{pmatrix}\cos \beta & \sin \beta\\- \sin \beta & \cos \beta
\end{pmatrix} \begin{pmatrix}
G^-\\\Delta_{13}\end{pmatrix} ,\quad
\tan \beta = \frac{\sqrt{15}}{5}.
\end{eqnarray}
The masses of $\Delta_0$ and $\Delta_{13}$ are given by
\begin{eqnarray}
m_{\Delta_0}^2 = m_{\Delta_{13}}^2 = \frac{5 \lambda_3}{4} {v'}^2 ~,
\end{eqnarray}
which are also degenerate. 
\begin{eqnarray}
\mathcal{L}_{M_{h, \Delta_2}} & = & - \frac{1}{2}  \begin{pmatrix}
h & \rho_2 \end{pmatrix}
\begin{pmatrix} 2 \lambda v^2 & \lambda_4 v v'\\ \lambda_4 v v'  & 2 \lambda'_2 {v'}^2 \end{pmatrix}
\begin{pmatrix}   h\\\rho_2\end{pmatrix} \nonumber\\
& = & - \frac{1}{2}  \left( \lambda v^2 + \lambda'_2 {v'}^2 - \sqrt{\left(
	\lambda v^2 - \lambda'_2 {v'}^2 \right)^2 + \lambda_4^2 v^2 v'^2} \right) h_1^2 \nonumber\\
&  & - \frac{1}{2}  \left( \lambda v^2 + \lambda'_2 {v'}^2 + \sqrt{\left(
	\lambda v^2 - \lambda'_2 {v'}^2 \right)^2 + \lambda_4^2 v^2 v'^2} \right) h_2^2, 
\end{eqnarray}
with
\begin{eqnarray}
\left( \begin{array}{c} h\\\rho_2\end{array} \right) 
= \left( \begin{array}{cc} \cos \alpha & \sin \alpha\\- \sin \alpha & \cos \alpha
\end{array} \right)
\left( \begin{array}{c} h_1\\h_2\end{array} \right),
\end{eqnarray}
where
\begin{eqnarray}
\tan \alpha =  \frac{\sqrt{\left( \lambda'_2 {v'}^2 - \lambda v^2\right)^2 + (\lambda_4 v v')^2} - \left( \lambda'_2 {v'}^2 - \lambda v^2\right)}{\lambda_4 v v'}.
\end{eqnarray}
If we identify $h_1$ as the $125 \si{GeV}$ SM-like Higgs field, then its quartic coupling and the mass of the heavier Higgs field $h_2$ are 
\begin{eqnarray}
\lambda \approx \frac{m_{h_1}^2}{2 v^2} + \frac{\lambda_4^2 {v'}^2}{2 \left( 2
	\lambda'_2 {v'}^2 - m_{h_1}^2 \right)}, \quad m_{h_2}^2  \approx  2 \lambda'_2 {v'}^2 + \frac{\lambda_4^2 v^2 {v'}^2}{2\lambda'_2 {v'}^2 - m_{h_1}^2}. 
\end{eqnarray}

In this case, both $X^\pm$ and $Z^\prime $ are dark matter candidates. There is a $Z_4 \times Z_2$ symmetry:
\begin{eqnarray}
Z_4 & : & \Delta_0 \rightarrow e^{i \pi} \Delta_0,
\quad \Delta_1 \rightarrow e^{i \frac{\pi}{2}} \Delta_1, \quad \Delta_2 \rightarrow \Delta_2,  \quad \Delta_{3}\rightarrow e^{-i \frac{\pi}{2}} \Delta_{3}, \notag\\
&  &  X^+ \rightarrow e^{-i\frac{\pi}{2}} X^+, \quad Z' \rightarrow Z'.\\
Z_2 & : & \Delta_0 \rightarrow \Delta_0 \quad \Delta_1 \rightarrow - \Delta_{- 1},
\quad \Delta_2 \rightarrow \Delta_{- 2}, \quad \Delta_3 \rightarrow -\Delta_{- 3}, \notag\\
&  & X^+ \rightarrow X^-, \quad Z' \rightarrow - Z'~,
\end{eqnarray}
which can prevent $X^\pm$ and $Z^\prime $ from decaying into SM particles.
Since we only focus on the situation that vector bosons are the only DM components, a condition:
\begin{eqnarray}
\lambda_3 \ge 64 g_4^2 /5~,\label{VDMcond2}
\end{eqnarray}
is required such that the new scalars $\Delta_0$ and $\Delta_{13}$ are massive enough to decay.

\section{Vacuum stability and perturbativity}
\label{sec:VSLP}
As the energy scale raises, all the couplings of interaction evolve according to renormalization group equations. Therefore, it is possible that some of the couplings change their sign at a critical scale. It means even the vacuum is stable in the low energy effective theory, it might become unstable in higher energy scale since the potential has no lower bound in the infinity of the vacuum configuration. In the SM, the quartic coupling of the Higgs field is an example that running to a negative value at the scale around $10^9$~GeV. It tells us that if the SM is valid up to Planck scale, then the vacuum configuration corresponding to the Higgs field is unstable or metastable. However, if there are some new physics appearing below the scale of switching sign, they can modify the running behaviors of the couplings, and thus the vacuum is capable to keep stable in high energy.

The co-positivity criteria~\cite{Kannike:2012pe} are used to ensure that the tree-level potential is bounded from below.
There are two cases which coincide with the categories of two inequivalent vacua, giving respectively~\cite{Cai:2015kpa}
\begin{eqnarray}
\lambda_3<0:
\begin{cases}
\lambda \ge 0,\\
\lambda'_2 + \frac{25}{32} \lambda_3 \ge 0,\\
\lambda_4 + 2\sqrt{\lambda (\lambda'_2+\frac{25}{32} \lambda_3)} \ge 0,
\end{cases}
\qquad
\lambda_3 \ge 0:
\begin{cases}
\lambda \ge 0,\\
\lambda'_2 \ge 0,\\
\lambda_4 + 2\sqrt{\lambda \lambda'_2} \ge 0.
\end{cases}
\label{eq:VS}
\end{eqnarray}
These conditions are required to be satisfied in the whole RG evolution processes till the cutoff scale.  
Another theoretical concern of the running couplings is their perturbativity in high energy. We expect them to keep finite below the Planck scale, otherwise they should be embedded in a more fundamental theory below that scale. The blowing up scale for the couplings is so called  Landau pole. 
The Landau pole problem of the $\SUtwoL$ scalar multiplets has been studied in Ref.~\cite{Hamada:2015bra}.
They showed that if the dimension of representation is larger than 4, the quartic couplings of the scalars blow up soon below the Planck scale even they are set to vanishing at the EW scale, and thus the same problem can be raised in our model.
However, since the gauge coupling of $\SUtwoD$ is not necessary to be the same as of $\SUtwoL$, the Landau pole scale can be totally different from the EW multiplet cases.
In order to determine the RG running of the couplings, we calculate their beta functions. To the one-loop level, the beta functions of couplings are shown as follows,
\begin{eqnarray}
\beta_{g_4} & = &  \frac{1}{16 \pi^2} \left(- \frac{8}{3} \right) g_4^3, 
\label{eq:betag4}\\
\beta_{\lambda'_2} & = & \frac{1}{16 \pi^2} \left( {30 \lambda'_2}^2 +
\frac{75}{8} \lambda^2_3 + 2 \lambda_4^2 + 15 \lambda'_2 \lambda_3 - 144
\lambda'_2 g_4^2 + 288 g_4^4 \right),  \label{eq:betal2p}\\
\beta_{\lambda_3} & = & \frac{1}{16 \pi^2} \left( \frac{63}{8} \lambda_3^2 +
24 \lambda'_2 \lambda_3 - 144 \lambda_3 g_4^2 + 288 g_4^4 \right), 
\label{eq:betal3}\\
\beta_{\lambda_4} & = & \frac{1}{16 \pi^2} \left( 4 \lambda_4^2 + 12 \lambda
\lambda_4 + 18 \lambda'_2 \lambda_4 + \frac{15}{2} \lambda_3 \lambda_4 -
\frac{9}{10} \lambda_4 g_1^2 - \frac{9}{2} \lambda_4 g_2^2 - 72 \lambda_4
g_4^2 + 6 \lambda_4 y_t^2 \right),  \label{eq:betal4}\\
\beta_{\lambda} & = & \beta_{\lambda}^{\text{SM}} + \frac{1}{16 \pi^2} 
\frac{7}{2} \lambda_4^2.  \label{eq:betal}
\end{eqnarray}
 where $g_1$ and $g_2$ are the gauge coupling coefficients of $\UoneY$ and $\SUtwoL$, $y_t$ is the top quark Yukawa coupling. Since their beta functions receive no modification from the dark sector, we refer their explicit expressions in the Ref.\cite{Cai:2015kpa}.

Before proceeding the numerical computation, it is useful to make some qualitative senses on the running behavior of each coupling.
The gauge coupling $g_4$ is assumed to be smaller than the gauge coupling of $\SUtwoL$, and thus its beta function $\beta_{g_4}$ is negligible implying $g_4$ approximately keep constant. 
Landau pole is inevitable to appear in $\beta_{\lambda'_2}$ and $\beta_{\lambda_3}$ if $g_4$ is not small enough. Note that there are $g_4^4$ terms in these two beta functions, which means even we turn off $\lambda'_2$ and $\lambda_3$ in low energy, they still increase fast owing to a nonzero $g_4$.
On the other hand, since all the terms in $\beta_{\lambda_4}$ has a factor $\lambda_4$, $\beta_{\lambda_4}$ has a fixed point at $\lambda_4=0$. It means the running behavior of $\lambda_4$ is under control if it is small at the scale of the dark sector. For achieving the DM freeze-out and vacuum stability, we need a small but nonzero $\lambda_4$. Note that the vacuum instability is usually brought out by $\lambda$ in the high energy due to a negative contribution of $y_t^4$ term in its beta function. The portal coupling $\lambda_4$ is helpful for easing the problem in two aspect. One is that the mixing between the SM Higgs doublet and the scalar septuplet leads to a "jump" of $\lambda$ around the mass scale of the septuplet due to the threshold effect~\cite{Elias-Miro:2012eoi}. The other one is the positive contribution of $\lambda_4^2$ in $\beta_\lambda$ can help to retard the decreasing of $\lambda$. To be precise, below the septuplet mass scale $\Lambda_\text{HS}$ which is assumed to about $\Lambda_\text{HS}=0.2v'$, septuplet decouples from the theory and thus the evolution of $\lambda$ is determined by the SM beta function of $\lambda$ (dropping the $\lambda_4^2$ term in Eq.\eqref{eq:betal}). At the scale $\Lambda_\text{HS}$, the threshold effect leads to the matching conditions as 
\begin{eqnarray}
\lambda(\Lambda_\text{HS}) = \lambda_\text{SM}(\Lambda_\text{HS}) + \frac{\lambda_4^2}{4\left( \lambda'_2 + \frac{25}{32}\lambda_3 \right)},\quad &&\text{for}~\lambda_3<0,\\
\lambda(\Lambda_\text{HS}) = \lambda_\text{SM}(\Lambda_\text{HS}) + \frac{\lambda_4^2}{4\lambda'_2}, \quad &&\text{for}~\lambda_3 \ge 0,
\end{eqnarray}
From the scale $\Lambda_\text{HS}$ toward higher energy, the $\lambda_4^2$ term is turned on in Eq.~\eqref{eq:betal}.

In our numerical analysis, the input values of the SM parameters at $Z$ pole applying the $\overline{\text{MS}}$ scheme are given in \cite{ParticleDataGroup:2020ssz,Hambye:1996wb}.
As for the couplings coefficients $g_4,\lambda'_2,\lambda_3$ and $\lambda_4$ are newly introduced, their input values at $\Lambda_\text{HS}$ are free parameters in our model.
Since the phenomenologies are mostly related to $\lambda_4$, while the Landau pole problem is related to $\lambda'_2$, we will mainly perform the experimental and theoretical constraints on the $\lambda'_2\text{-}\lambda_4$ plane. For the other couplings like $g_4$, $\lambda_3$, and $v'\sim\si{TeV}$, we select some typical values and then runs the RGE to determine the evolutions of all couplings toward high energy scale. If the vacuum stability conditions in Eq.~\eqref{eq:VS} are violated at any energy scale below the Planck scale, then the corresponding parameter set are marked down as unstable vacuum region. On the other hand, if any coupling excesses $4\pi$ below Planck scale, it becomes non-perturbative and meets Landau pole soon.

The numerical results show that the energy scale of Landau pole decreases as $|\lambda_3|$ growing up, and have a maximum at $g_4(\Lambda_\text{HS})\sim$ $0.1$--$0.3$ when other parameters are fixed. However, as we are focusing on the case that the vector bosons are the only stable DM candidates, we will further assume some relations between $g_4^2$ and $\lambda_3$ (see Eq.\eqref{VDMcond1} and \eqref{VDMcond2}).
In those cases, we find that the perturbative region of parameters is enlarged as $g_4$ decreases.
The upper limit for the absolute value of the input portal coupling is found to be $|\lambda_4(\Lambda_\text{HS})| \lesssim 0.24$, while the upper limit of the input gauge coupling is found to be $|g_4(\Lambda_\text{HS})| \lesssim 0.15$ for $\lambda_3(\Lambda_\text{HS})\ge0$, or $g_4(\Lambda_\text{HS}) \lesssim 0.19$ for $\lambda_3(\Lambda_\text{HS})<0$.

\begin{figure}[!t]
	\centering	
	{\includegraphics[width=0.5\textwidth]{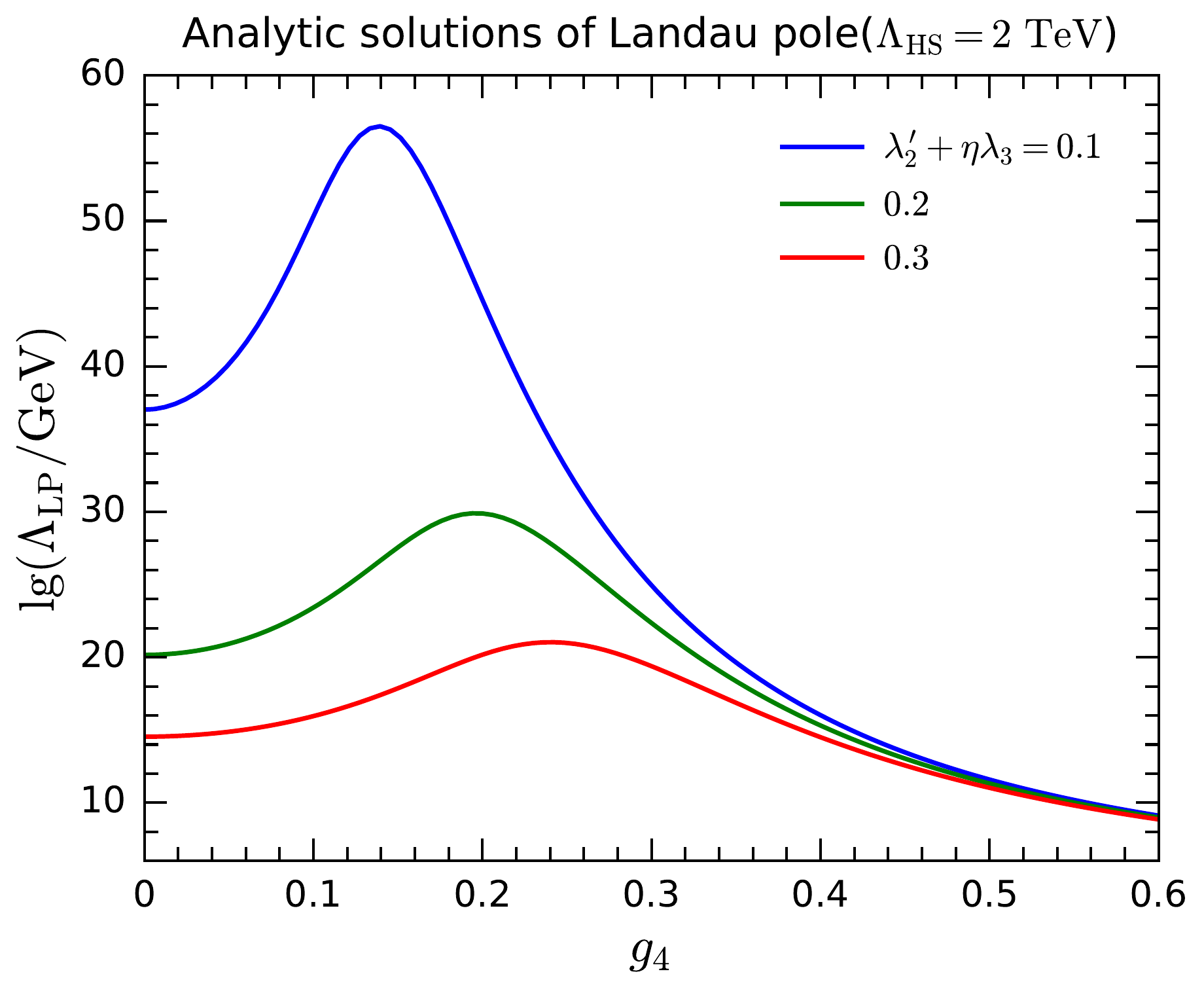}}
	\caption{The analytic solutions of Landau pole as a function of $g_4(\Lambda_\text{HS})$ with different values of $\lambda'_2+\eta \lambda_3$.
		The blue, green and red lines are the solutions at $\lambda'_2+\eta \lambda_3=0.1,0.2$ and $0.3$ respectively, with $\eta \approx 0.9$.}
	\label{fig:AnalyticSolution}
\end{figure}

Naively thinking, the Landau pole may decrease monotonically as $g_4$ growing.
However, an analytical treatment of solving the RGEs in Eq.~\eqref{eq:LP_analy} shows that if $\lambda'_2+\eta \lambda_3 >0$ (see more details in Appendix~\ref{sec:append}), the Landau pole scale is not a monotonic function of $g_4(\Lambda_\text{HS})$. For a given set of other parameters, the Landau pole scale $\Lambda_{\text{LP}}$ reaches its maximal value at a non-zero value of $g_4(\Lambda_\text{HS})$ (see Fig.~\ref{fig:AnalyticSolution}).
This behavior is also confirmed by the full numerical calculations.
We can utilize the approximate analytical solution to estimate the upper limit of $g_4(\Lambda_\text{HS})$ under the scalar decay conditions.
For $\lambda_3 \ge 0$, the analytic estimation gives $g_4(\Lambda_\text{HS}) \lesssim 0.15$ which agrees with the numerical calculation very well.
For $\lambda_3 < 0$, it gives $g_4(\Lambda_\text{HS}) \lesssim 0.37$ which deviates from the numerical calculation by a factor of 2.

\section{Constraining the model by phenomenologies, perturbativity, and vacuum stability}
\label{sec:DM}

In this section, we study the phenomenological constraints for the vector bosons DM, associated with the theoretical constraints from perturbativity and vacuum stability.
Similar to most of the the other cold dark matter models, the vector DMs in our model are produced via the freeze-out mechanism. We utilize the \texttt{FeynRules~2}~\cite{Alloul:2013bka} and \texttt{micrOMEGAs}~\cite{Belanger:2014vza}  packages to determine the relic density and annihilation cross section of DMs. We compare the numerical result of DM relic density with the experimental result, $\Omega_\text{CDM} h^2 = 0.12$, from the Planck experiment~\cite{Planck:2018vyg}.
We also consider the indirect detection bounds on DM annihilation cross section provided by Fermi-LAT~\cite{MAGIC:2016xys}, which combines the 6 years data of gamma-ray observations from Segue 1 with 158 hours and 15 dwarf satellite galaxies. In our numerical calculation of the annihilation cross section, DMs in the dwarf satellite galaxies are assumed to be at rest for simplicity. The future prospect on the sensitivity of DM indirect detection from the Cherenkov Telescope Array (CTA)~\cite{CTAConsortium:2012fwj} is included for comparison.
In addition, we consider the most recent direct detection constraints provided by first result of LUX-ZEPLIN (LZ) experiment~\cite{LUX-ZEPLIN:2022qhg}, and its prospective sensitivity of 1000 days data~\cite{LZ:2015kxe}. 
Apart from the DM phenomenologies, there are LHC bounds on the the mixing angle between the Higgs field and the scalar $\rho_{2,3}$. In our model, the trilinear coupling between Higgs-like field and all other SM fields are suppressed by a factor $\cos\alpha$ due to the mixing. We fit the measurements of couplings $\kappa_{i},~i=\gamma,W,Z,g,t,b,\tau$ from ATLAS~\cite{ATLAS:2019nkf} and CMS~\cite{CMS:2018uag} with a common $\kappa$, and find $\kappa>0.9875$ at 95\% CL. We will set this bound for $\kappa=\cos\alpha$ in our model.

In the theoretical aspect, we assume that there are no other new sectors directly couple to the SM and dark sector below the Planck scale, and thus the perturbativity and vacuum stability conditions should be satisfied below the Planck scale. The RGEs are solved numerically with certain input of coupling parameters at the mass scale of the septuplet (around TeV).

\subsection{Splitting case $(\lambda_3<0)$}

\begin{figure}[!t]
	\centering
	\subfigure[~DM annihilate to $W$ and $Z$. \label{fig:anni_WZ}]
	{\includegraphics[width=0.5\textwidth]{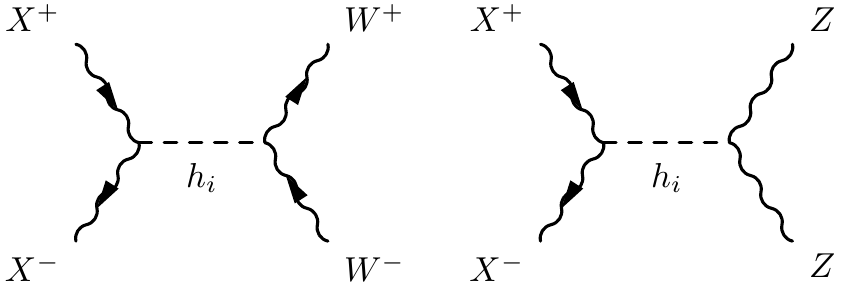}}
	\subfigure[~DM annihilate to SM-like Higgs $h_1$.]
	{\includegraphics[width=1\textwidth]{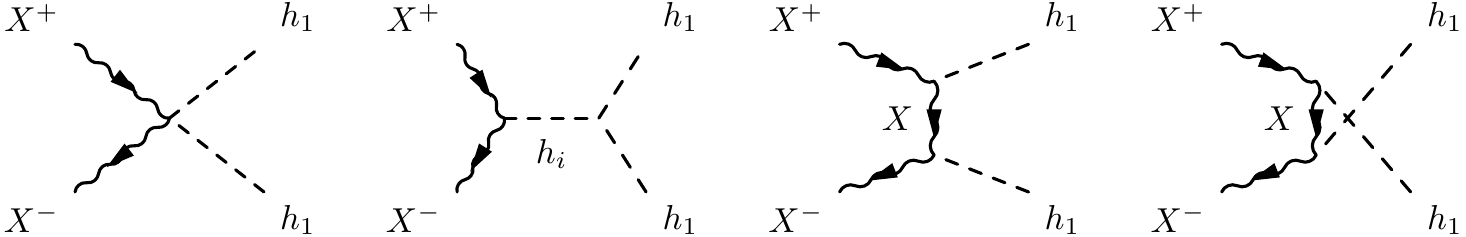}}
	\caption{The dominant annihilation channels of DM.}
	\label{fig:FeynDiagram}
\end{figure}

Since $X^\pm$ couples to the SM particles through Higgs-portal, their annihilation cross section at freeze-out epoch closely relate to the mixing angle $\alpha$ between $h$ and $\rho_3$.
The dominant DM annihilation channels are $X^++X^-\to W^++W^-,~Z+Z,~h_1+h_1$, as displayed in Fig.~\ref{fig:FeynDiagram}. The thermal averaged annihilation cross section corresponding to these processes are given by
\begin{eqnarray}
\sum_{ij=WW,ZZ,h_1h_1}\langle\sigma_{X^+X^-\to ij}v\rangle\approx\frac{\sin ^{2} (2 \alpha)m_{X}^{2}}{48 \pi v^{2} v^{\prime 2}} \left(1-\frac{4m_X^2}{m_{h_2}^2}\right)^{-2}
\end{eqnarray}
in the $m_{h_2},m_X\gg m_{W},m_Z,m_{h_1}$ limit. This approximately yields a relic density of DM:
\begin{eqnarray}
\Omega_{\text{DM}}h^2\approx 2.59\times\left(\frac{0.1}{g_4}\right)^2\left(\frac{0.1}{\sin^2(2\alpha)}\right)\left(1-\frac{4m_X^2}{m_{h_2}^2}\right)^{2}.
\end{eqnarray}
Since the Higgs couplings measurements imply a $\sin^2(2\alpha)\lesssim0.1$, we need a $g_4(\Lambda_\text{HS})\approx0.4$ with $4m_X^2\ll m_{h_2}^2$ or a $g_4(\Lambda_\text{HS})<0.4$ with $4m_X^2\approx m_{h_2}^2$ (closed to the resonance) in order to obtain an $\Omega_{\text{DM}}h^2\approx0.12$. As we have estimated in the previous section, there is an upper bound for $g_4(\Lambda_\text{HS})$ from the perturbativity as $g_4(\Lambda_\text{HS})\lesssim0.19$, and thus the correct relic density can only be achieved near the resonance.

\begin{figure}[!t]
	\centering
	\subfigure[~$g_4=0.05,~\lambda_3=-0.2,~v'=10~\si{TeV}$.\label{fig:DMLPVSv3_1}]
	{\includegraphics[width=0.49\textwidth]{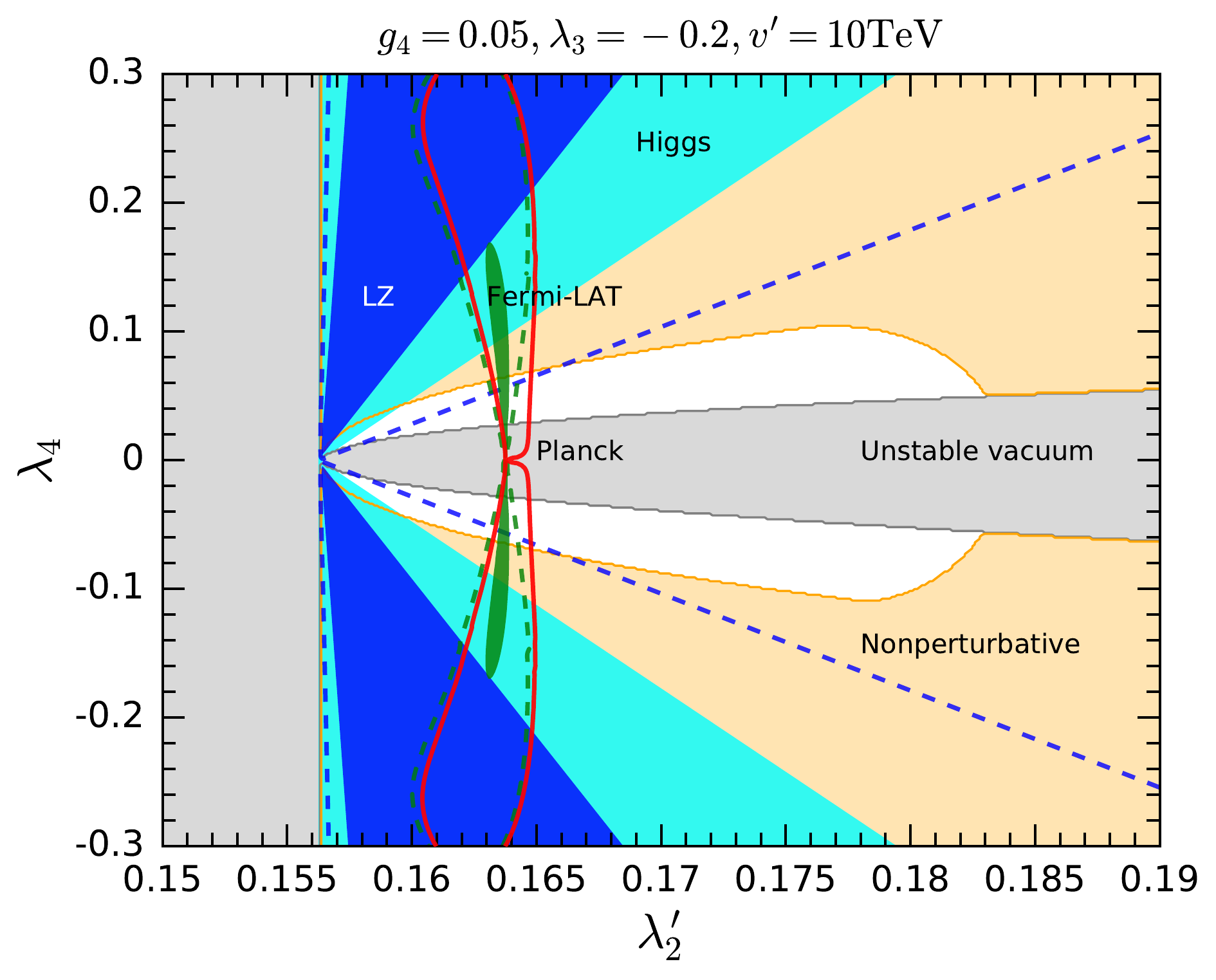}}
	\subfigure[~$g_4=0.1,~\lambda_3=-0.2,~v'=10~\si{TeV}$.\label{fig:DMLPVSv3_2}]
	{\includegraphics[width=0.49\textwidth]{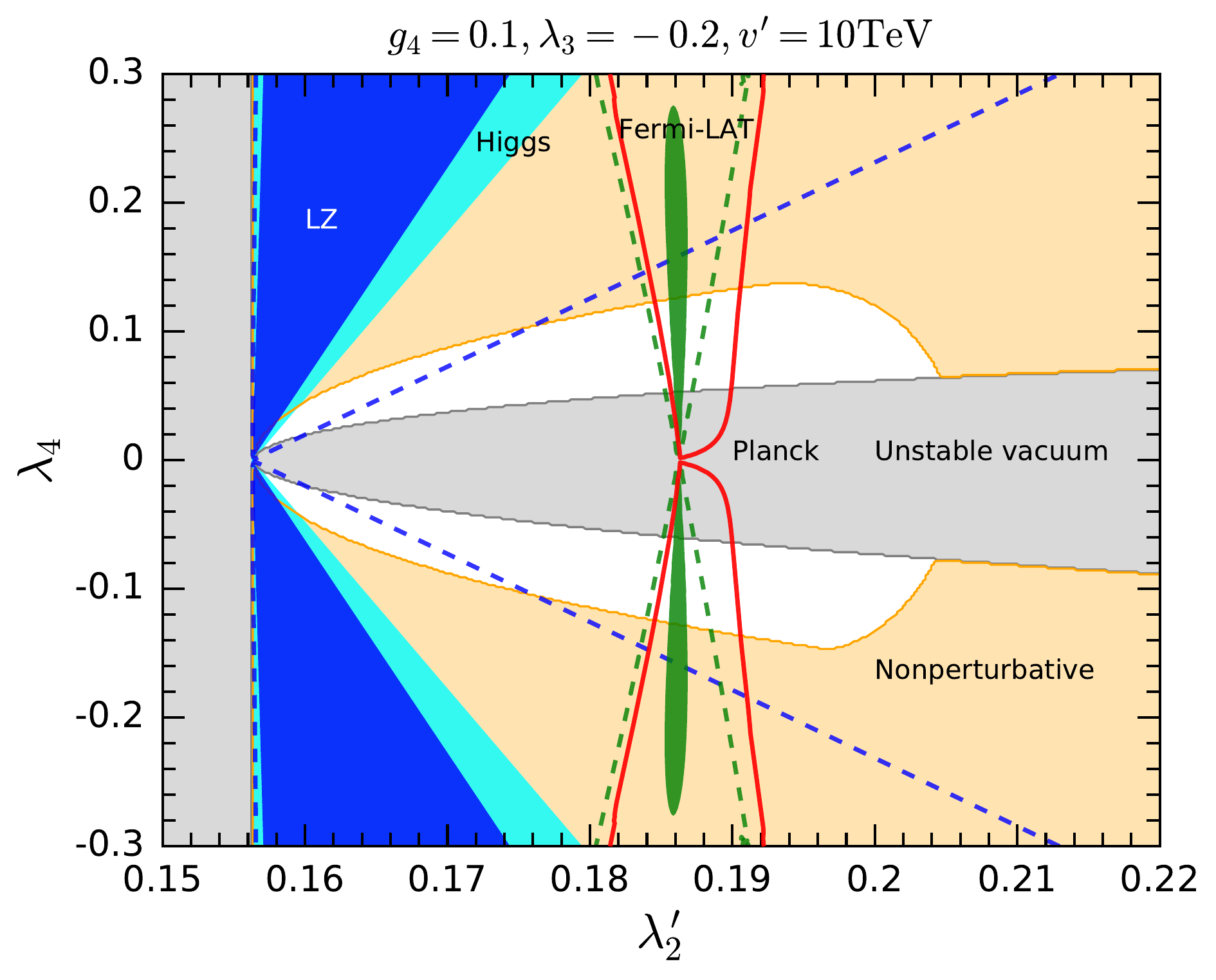}}
	\caption{Splitting case: constraints on $\lambda'_2$-$\lambda_4$ plane from perturbativity, vacuum stability and experiments including DM relic density, direct and indirect detection, and LHC bounds on Higgs couplings.
		The VEV of the septuplet $v'$ is fixed as $10~\si{TeV}$, while $\{g_4,\lambda_3\}$ are set to be $\{0.05,-0.2\}$ (left panel) and $\{0.1,-0.2\}$ (right panel), respectively.
		The gray regions correspond to parameter points that violating the vacuum stability conditions given in Eq.~\eqref{eq:VS}, while the orange regions correspond to the running couplings becoming non-perturbative below the Planck scale.
		The regions shaded in blue are excluded by the upper limit on DM-nucleon scattering cross section from LZ $90\%$ C.L.~\cite{LUX-ZEPLIN:2022qhg}, while the regions shaded in green are excluded by the upper limit on $\svann$ in the $W^\pm$ channel from Fermi-LAT at $95\%$ C.L.~\cite{MAGIC:2016xys}.
		The blue dashed lines are the sensitivity curves from the prospect of LZ with 1000-days data~\cite{LZ:2015kxe}, while the green dashed lines are the sensitivity of CTA~\cite{CTAConsortium:2012fwj}.
		The red lines indicate a DM relic density value $\Omega_\text{DM}h^2=0.12$~\cite{Planck:2018vyg}. The region shaded in cyan are excluded by the Higgs couplings measurements in LHC~\cite{ATLAS:2019nkf,CMS:2018uag}}
	\label{fig:DMLPVSv3}
\end{figure}

In Fig.~\ref{fig:DMLPVSv3}, we show the perturbativity, vacuum stability and DM phenomenological constraints in the $\lambda'_2$-$\lambda_4$ plane, with other parameters fixed. The DM mass in Fig.~\ref{fig:DMLPVSv3_1} and Fig.~\ref{fig:DMLPVSv3_2} are $612.4~\si{GeV}$ and $1.225~\si{TeV}$, respectively.
The regions shaded in gray are excluded since they lead to an unstable vacuum. Note that, these regions separate in two pieces. One is in the area that $\lambda'_2 + 25\lambda_3/32<0$, which violates the second condition given in the left part of Eq.\eqref{eq:VS}. Another area of exclusion (bullet shape) is mainly caused by $\lambda(\mu)<0$ at some higher energy scale (the first condition in the left part of Eq.\eqref{eq:VS}).
We can see that a small but non-vanishing $\lambda_4$ is needed for preventing $\lambda(\mu)$ from running into negative value. The boundary of the bullet shape area can be estimated by $\lambda_4^2/(\lambda'_2+25\lambda_3/32)\approx0.1$, which is given by the increment of $\lambda$ due to the threshold effect.
As we have mentioned previously, high dimensional representation of $\SUtwoD$ leads to a large positive contribution in the beta functions, and thus leads to stringent constraints of perturbativity for the parameters. The regions shaded in orange are excluded due to some coupling becoming non-perturbative ($>4\pi$) in high energy scale. The white region in the plots satisfy both the perturbativity and vacuum stability conditions. Numerical solutions of RGEs show that the allowed regions (white) shrink when $|\lambda_3(\Lambda_\text{HS})|$ increases.
Note that although the pattern in Fig.~\ref{fig:DMLPVSv3} looks almost symmetric about the $\lambda_4=0$ axis, there is a slight asymmetry due to the $\lambda_4^2$ term in $\beta_{\lambda_4}$ (see Eq.~\eqref{eq:betal4}).
We can also see that the allowed region in Fig.~\ref{fig:DMLPVSv3_1} with $g_4=0.05$ is smaller than the allowed region in Fig.~\ref{fig:DMLPVSv3_2} with $g_4=0.1$, which qualitatively agrees with our previous analytical analysis. The Landau pole scale is not monotonically depending on the $g_4$, and thus a smaller $g_4$ does not always mean a better situation for keeping perturbative in high energy.

In the DM phenomenological aspect, the DM observables mainly rely on $\{v',g_4,\lambda'_2+25\lambda_3/32,\lambda_4\}$ at tree level. The red contours in Fig.~\ref{fig:DMLPVSv3} corresponds to DM relic density $\Omega_{\text{DM}} h^2\approx0.12$, while the region shaded in green and blue are excluded by current DM indirect and direct detection experiments, respectively. The green and blue dashed lines show the future prospect from the CTA and LZ sensitivities. The region shaded in cyan are excluded by the LHC measurement on the Higgs couplings.
Note that the red contour concentrates in a narrow region of $\lambda'_2$ since the correct relic density is obtained around the resonance point of the internal propagators. It can be used to estimate the value of $\lambda'_2$ if all other parameters are input. To be precise, the resonance peak appears at the parameter point such that $m_{h_2}\approx 2m_{X^\pm}$, which gives
\begin{eqnarray}
	\lambda'_2+\frac{25}{32}\lambda_3 \approx \frac{3}{2}g_4^2 + \sqrt{\left(\frac{3}{2}g_4^2\right)^2 - \frac{\lambda_4^2 v^2}{4 v'^2}},
\end{eqnarray}
where we have dropped the $m_{h_1}$ since it is negligible comparing to the dark sector masses.

For the direct detection, the spin-independent cross section at tree level between DM and nucleon $N = p, n$ is given by~\cite{Hambye:2008bq}
\begin{eqnarray}
	\sigma_{\text{SI}} = \frac{f^2 m_N^4}{4 \pi (m_{X} + m_N)^2} \frac{\sin^2 (2 \alpha) m_{X}^4 }{v^2 {v'}^2} \frac{(m_{h_2}^2 - m_{h_1}^2)^2}{m_{h_1}^4
	m_{h_2}^4},
\end{eqnarray}
where $f\equiv\langle N|\sum_{q} m_q \bar{q} q|N \rangle$.
Since $m_{h_2} \gg m_{h_1}$, we can eliminate $m_{h_2}$ in the numerator and denominator in a rough estimation.
On the other hand, from Eq.~\eqref{eq:alpha_v3}, the mixing angle $\alpha$ approximately depends on $(\lambda'_2+25\lambda_3/32)$ and $\lambda_4$ as follows
\begin{eqnarray}
	|\sin\alpha|\approx|\tan\alpha| \approx \sqrt{\left(\frac{\lambda'_2+\frac{25}{32}\lambda_3}{|\lambda_4|}\right)^2\frac{v'^2}{v^2}+1} - \frac{\lambda'_2+\frac{25}{32}\lambda_3}{|\lambda_4|}\frac{v'}{v}\approx \frac{|\lambda_4|}{2(\lambda'_2+\frac{25}{32}\lambda_3)}\frac{v}{v'},
\end{eqnarray}
where $\lambda v^2$ has been neglected when compare with the large $(\lambda'_2+25\lambda_3/32)v'^2$. Therefore, we can estimate
\begin{eqnarray}
\sigma_{\text{SI}} \approx \left(\frac{f^2 }{4 \pi }\right)\left(\frac{m_N^4}{m_{h_1}^4}\right)\left(\frac{m_{X}^2}{v'^4}\right)\frac{|\lambda_4|^2}{(\lambda'_2+\frac{25}{32}\lambda_3)^2}.
\end{eqnarray}
If $m_X,~g_4,~v'$ are fixed, then the combination $|\lambda_4|/(\lambda'_2+25\lambda_3/32)$ completely determines the cross section $\sigma_{\text{SI}}$.
Consequently, the boundaries of blue region are nearly straight lines. The bounds from the Higgs couplings measurements have similar behaviors since they only depends on the mixing angle $\alpha$.
In both panels of Fig.~\ref{fig:DMLPVSv3}, there are regions consistent with all the current experimental and theoretical constraints. We find that the $\lambda'_2$ is almost pinned down in a narrow region by the DM indirect detection and relic density, when $g_4$, $\lambda_3$ and $v'$ are fixed. On the other hand, perturbativity and vacuum stability conditions are available to constrain the $|\lambda_4|$. Current direct detection is less effective as the perturbativity bound, however the sensitivity of LZ with 1000-days data might be powerful enough to probe the survival area. The Higgs couplings bounds are usually stronger than current direct detection bounds, but still not effective enough to probe the region allowed by perturbativity. 
From the prospect of the CTA experiment, we find it has potential to confirm or exclude a part of the contour implied by the relic density.

\subsection{Degenerate case $(\lambda_3 \ge 0)$}

In this case, both the complex vector field $X^\pm $ and the real vector $Z'$ are DM candidates.The dominant annihilation processes for the $X^\pm$ are the same as in Fig.~\ref{fig:FeynDiagram}. For the $Z'Z'$ annihilation, we have similar diagrams and amplitudes (a factor $1/2$ appears since it only has a half degree of freedom of $X^\pm$).
The semi-annihilations like $X^+ X^-\to Z' h_1$ and $X^\pm Z' \to X^\pm h_1$ channels and other processes are sub-dominant. The annihilation cross section of DM are similar to the previous case, and then the relic density of DM is
\begin{eqnarray}
\Omega_{\text{DM}}h^2\approx 1.46\times\left(\frac{0.1}{g_4}\right)^2\left(\frac{0.1}{\sin^2(2\alpha)}\right)\left(1-\frac{4m_X^2}{m_{h_2}^2}\right)^{2}.
\end{eqnarray}
Note that there is an upper bound for the gauge coupling, $g_4(\Lambda_\text{HS})\lesssim0.15$, required by the perturbativity, and an upper bound for $\sin^2(2\alpha)$, $\sin^2(2\alpha)\lesssim0.1$, required by the Higgs couplings measurements. Therefore, in order to achieve observed relic density of DM, $\Omega_{\text{DM}}h^2\approx0.12$, the masses of $X^\pm,Z'$ and $h_2$ should be near the resonance $2m_{X,Z'}\approx m_{h_2}$: 
\begin{eqnarray}
\lambda'_2 \approx 4 g_4^2 + \sqrt{\left(4 g_4^2\right)^2 - \frac{\lambda_4^2 v^2}{4 v'^2}}.
\end{eqnarray}

\begin{figure}[!t]
	\centering
	\subfigure[~$g_4=0.1,~\lambda_3=0.13,~v'=10~\si{TeV}$.\label{fig:DMLPVSv2_1}]
	{\includegraphics[width=0.49\textwidth]{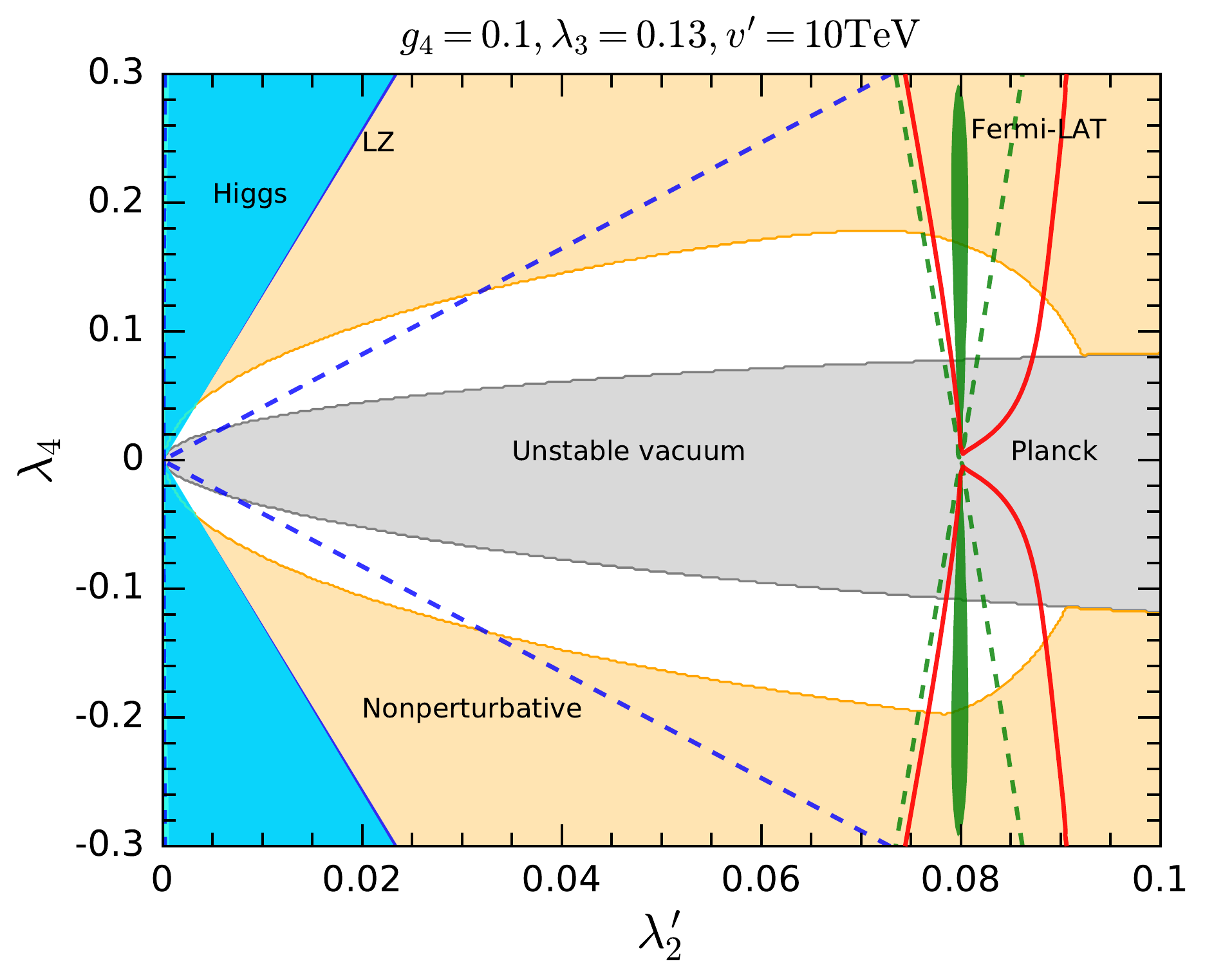}}
	\subfigure[~$g_4=0.1,~\lambda_3=0.13,~v'=5~\si{TeV}$.\label{fig:DMLPVSv2_2}]
	{\includegraphics[width=0.49\textwidth]{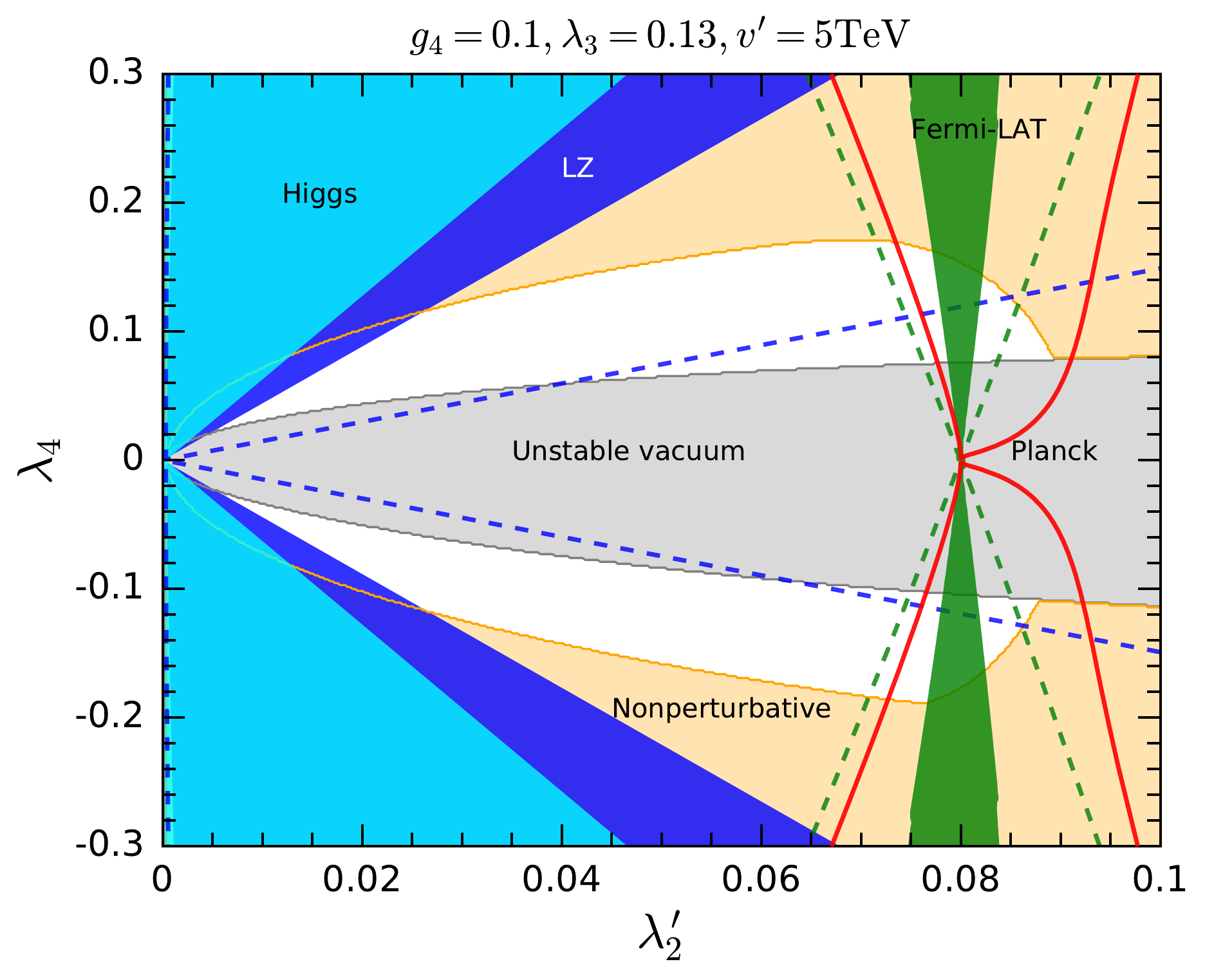}}
	\subfigure[~$g_4=0.05,~\lambda_3=0.2,~v'=10~\si{TeV}$.\label{fig:DMLPVSv2_3}]
	{\includegraphics[width=0.49\textwidth]{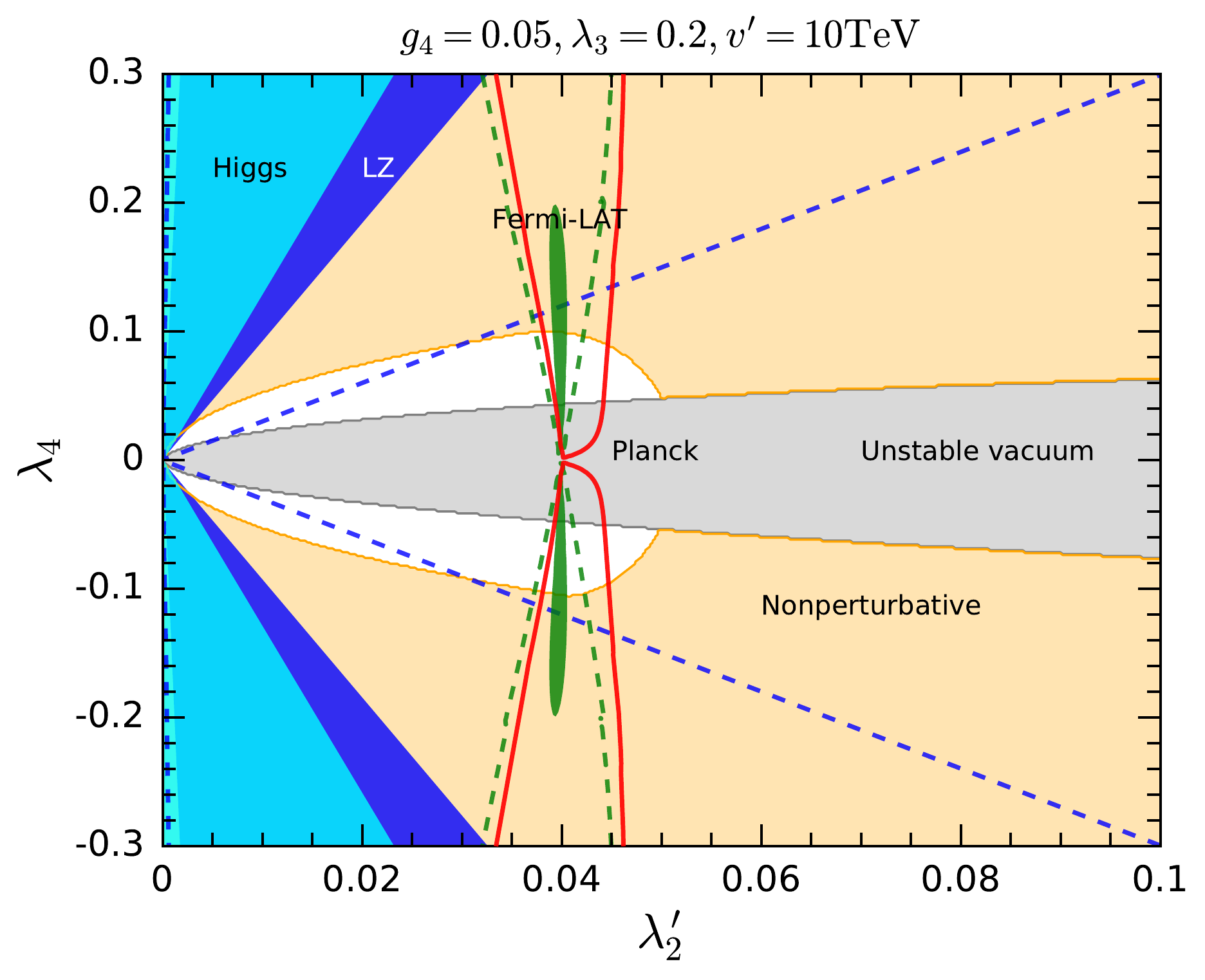}}
	\caption{Degenerate case: constraints in $\lambda'_2$-$\lambda_4$ parameter plane by vacuum stablity, perturbativity and DM experiments including relic density, direct and indirect detection.
	The parameters $\{g_4,\lambda_3,v'\}$ are set to be $\{0.1,0.13,10~\si{TeV}\}$ (top left panel), $\{0.1,0.13,5~\si{TeV}\}$ (top right panel) and $\{0.05,0.2,10~\si{TeV}\}$ (bottom panel), respectively.
		The meanings of different colors are the same as Fig.~\ref{fig:DMLPVSv3}.}
	\label{fig:DMLPVSv2}
\end{figure}

The experimental and theoretical constraints on the $\lambda'_2$-$\lambda_4$ plane for this case are presented in Fig.~\ref{fig:DMLPVSv2}.
The DM masses in Fig.~\ref{fig:DMLPVSv2_1},~\ref{fig:DMLPVSv2_2} and \ref{fig:DMLPVSv2_3} are set to $2~\si{TeV}$, $1~\si{TeV}$ and $1~\si{TeV}$ respectively.
Note that the DM observables in this case only rely on $(v',g_4,\lambda'_2,\lambda_4)$, without $\lambda_3$ dependency. 
The spin independent direct detection cross section and Higgs couplings depend on the mixing angle given by
\begin{eqnarray}
|\tan \alpha| \approx \sqrt{\left(\frac{\lambda'_2}{|\lambda_4|}\right)^2\frac{v'^2}{v^2}+1} - \frac{\lambda'_2}{|\lambda_4|}\frac{v'}{v}\approx \frac{|\lambda_4|v}{2\lambda'_2 v'},
\end{eqnarray}
and thus their bounds are straight lines on the $\lambda'_2$-$\lambda_4$ plane. Note that in Fig.~\ref{fig:DMLPVSv2_1}, the Higgs coupling bound nearly coincides with the direct detection bound, and thus the blue area is nearly covered by the cyan area. 

Varying $v'$ from $\si{TeV}$ to $10~\si{TeV}$ does not change the perturbativity and vacuum stability bounds significantly.
However, both the direct and indirect detection experiment bounds are sensitive to $v'$. By comparing Fig.~\ref{fig:DMLPVSv2_1} and Fig.~\ref{fig:DMLPVSv2_2}, we can see that the excluded area are enlarged when $v'$ decreases.
Finally, we find that both prospective sensitivities of LZ and CTA experiments can probe all the available region shown in Fig.~\ref{fig:DMLPVSv2_2}.

\section{Conclusion and discussion}
\label{sec:conclusion}

The vector dark matter originates from a hidden gauged $\SUtwoD$ is investigated in this work. The $\SUtwoD$ is spontaneously broken by the vacuum configuration of a real scalar septuplet. The hidden dark sector communicate with the SM via the Higgs portal, and thus all the phenomenologies are related to the mixing angle between the SM Higgs and the radial component of the septuplet. 
We find that there are two different vacuum configurations corresponding to different sign of the quartic coupling $\lambda_3$. 
For the case of $\lambda_3<0$, the mass spectrum of gauge fields and scalar fields are splitting, and the lightest particle is assumed to be the complex vactor field $X^\pm$. We find a remnant discrete symmetry $Z_6 \times Z_2$ after the SSB, which can prevent the DM candidate from decaying into the SM particles.
For the case of $\lambda_3\ge0$, the mass spectrum of gauge fields are degenerate. Both the complex vector $X^\pm$ and the real vector $Z'$ are dark matter candidates. A remnant discrete symmetry $Z_4 \times Z_2$ is found in this case, and thus the DM candidates are stable.

We also calculate the one-loop beta functions of couplings and let the couplings evolve to higher energy scale.
We find that $\lambda'_2$ and $\lambda_3$ will soon blow up in higher energy if $g_4$, $\lambda'_2$ and $\lambda_3$ are not small enough. If we assume that all the couplings should keep perturbative below Planck energy scale, only a very restricted parameter space can survive.
Generally, the scale of Landau pole decreases when $|\lambda_3|$ grows. On the other hand, the scale of Landau pole is not monotonically dependent on $g_4$. We find that the scale of Landau pole reaches its maximum around $g_4\sim0.1$--$0.3$, which is confirmed by both analytical and numerical calculations.
However, if we assume that all the dark sector scalar can completely decay into the vector dark matter, which corresponds to $\lambda_3 \le -48 g_4^2/5$ for $\lambda_3<0$ and $\lambda_3 \ge 64 g_4^2/5$ for $\lambda_3 \ge 0$, then the perturbative parameter regions always shrink when $g_4$ increases in the region of our interests.
As a result, there are upper limits for the absolute value of the portal coupling, $|\lambda_4| \lesssim 0.24$, and the gauge coupling $g_4 \lesssim 0.15~(0.19)$ for the $\lambda_3\ge0~(<0)$ case.

The range of the DM masses is from several hundreds of $\si{GeV}$ to a few $\si{TeV}$.
To achieve the observed relic density, the masses of DM and the heavy Higgs should be closed to the resonance region, $2m_X\approx m_{h_2}$ in both cases.
We also find that current direct and indirect detection of DM and the LHC measurements of Higgs couplings can barely exclude the region recommended by the DM relic density and perturbativity, unless the DM mass is around hundreds of $\si{GeV}$.
However, these regions can be tested by the future searches in the LZ and CTA experiments.

In this work, we only consider the case that vector fields constitute all the components of DM. However, some scalar particles in the dark sector can be parts of the DM as well if they are not massive enough for decaying into vector particles. In this case, the co-annihilation between the vectors and scalars can play a considerable role during the DM freeze-out.
We will leave this more complicated situation as a future research.

\begin{acknowledgments}
We thank Zhao-Huan Yu for helpful discussions. This work is supported by the National Natural Science Foundation of China (NSFC) under Grant Nos. 11875327 and 11905300, the Fundamental Research Funds for the Central Universities, the Natural Science Foundation of Guangdong Province, and the Sun Yat-Sen University Science Foundation.

\end{acknowledgments}

\appendix
\section{Analytic solution of Landau pole}
\label{sec:append}

The one-loop beta functions of gauge couplings $\beta_{g_i} =
\frac{b_i}{16 \pi^2} g_i^3$ can be solved as,
\begin{eqnarray}
g_i (\mu) = \left( \frac{1}{g_i^2 (\Lambda_0)} - \frac{b_i}{8 \pi^2} \ln
\frac{\mu}{\Lambda_0} \right)^{- \frac{1}{2}} .
\end{eqnarray}
At the one-loop level, gauge and Yukawa couplings are the same as the SM and they do not meet Landau pole below the Planck scale.

The couplings $\lambda'_2$ or $\lambda_3$ are easy to blow up in high energy even they are set to vanish in low energy scale. Their beta functions are given by 
\begin{eqnarray}
\frac{d \lambda'_2}{d t'} & = & {30 \lambda'_2}^2 + \frac{75}{8} \lambda^2_3
+ 15 \lambda'_2 \lambda_3 - 144 \lambda'_2 g_4^2 + 288 g_4^4, 
\label{eq:betal2p2}\\
\frac{d \lambda_3}{d t'} & = & \frac{63}{8} \lambda_3^2 + 24 \lambda'_2
\lambda_3 - 144 \lambda_3 g_4^2 + 288 g_4^4 .  \label{eq:betal32}
\end{eqnarray}
where $t' \equiv \frac{\ln \mu}{16 \pi^2}$. We have assumed that $\lambda_4$ is small and the $\lambda_4^2$ term are neglected for simplicity.
To solve these differential equations, we define functions $f_1 (t')$ and $f_2 (t')$ as follows,
\begin{eqnarray}
\lambda'_2 (t') = f_1 (t') g_4^2 (t'), \qquad \lambda_3 (t') = f_2 (t')
g_4^2 (t'),
\end{eqnarray}
and then replace them to above two beta functions:
\begin{eqnarray}
\frac{d f_1}{d G} & = & 30 f_1^2 + \frac{75}{8} f_2^2 + 15 f_1 f_2 - (144 +
2 b_4) f_1 + 288, \\
\frac{d f_2}{d G} & = & \frac{63}{8} f_2^2 + 24 f_1 f_2 - (144 + 2 b_4) f_2
+ 288, 
\end{eqnarray}
where $G(t')$ is given by
\begin{eqnarray}
d G \equiv g_4^2 d t' = \frac{1}{b_4 g_4} \frac{d g_4}{d t'} d t' \quad &
\Rightarrow & \quad G (t') = \frac{1}{b_4} \ln \frac{g_4 (t')}{g_4 (t'_0)} .
\end{eqnarray}

The next step is to introduce another function, $F = f_1 + \eta f_2$, and adjust $\eta$ to eliminate the $F f_1$ term in
\begin{eqnarray}
\frac{d F}{d G} & = & \left( \frac{75}{8 \eta^2} + \frac{63}{8 \eta} \right) F^2 - (144 + 2
b_4) F + 288 (1 + \eta) + \left( \frac{75}{8 \eta^2} - \frac{57}{8 \eta} + 6
\right) f_1^2 \nonumber\\
&  & - \left( \frac{75}{4 \eta^2} + \frac{3}{4 \eta} - 24 \right) F f_1. 
\end{eqnarray}
There are two roots for the equation, $75/(4 \eta^2) + 3/(4 \eta) - 24 =0$, and $\eta = \left( 1 + \sqrt{3201} \right) / 64 \approx 0.899647$ is chosen since it gives a smaller coefficient of $f_1^2$ term.
As a result, the evolution equation of $F$ can be written as
\begin{eqnarray}
\label{eq:dF}
\frac{d F}{d G} = c_2 F^2 - c_1 F + c_0 + c' f_1^2,
\end{eqnarray}
where
\[ c_2 = \frac{24 \left( 1621 + 21 \sqrt{3201} \right)}{\left( \sqrt{3201} + 1
	\right)^2} \approx 20.3366, \quad c_1 = 144 + 2 b_4 \approx 138.667, \]
\begin{eqnarray}
\quad c_0 = \frac{9 \left( 65 + \sqrt{3201} \right)}{2} = 547.098, \quad c' = \frac{6 \left( 9526 - 74
	\sqrt{3201} \right)}{\left( \sqrt{3201} + 1 \right)^2} = 9.66339.
\end{eqnarray}
If we ignore the $f_1^2$ term and $F$ can be solved analytically
\begin{eqnarray}
F (G) = \frac{c_1}{2 c_2} + \hat{d} \tan \left\{ c_2  \hat{d} G (t') +
\tan^{- 1} \left[ \frac{1}{\hat{d}} \left( F (t'_0) - \frac{c_1}{2 c_2}
\right) \right] \right\},
\end{eqnarray}
with $\hat{d} \equiv \sqrt{\frac{c_0}{c_2} - \frac{c_1^2}{4 c_2^2}}
\approx$5.1543.
When $\lambda'_2$ or $\lambda_3$ (equivalently $f_1$ or
$f_2$), meets Landau pole at some scale $\Lambda_{\text{LP}}$, $F$ also blows up.
Therefore we can extract the Landau pole $\Lambda_{\text{LP}}$ in this model by analyzing the divergence of $F$.
Since $G(t')$ increases monotonically with $t'$ and $G (t'_0)= 0$, we can find $F (t')$ diverging at
\begin{eqnarray}
c_2  \hat{d} G (t') + \tan^{- 1} \left[ \frac{1}{\hat{d}} \left( F (t'_0) -
\frac{c_1}{2 c_2} \right) \right] = \frac{\pi}{2}.
\end{eqnarray}
In other words, the Landau pole is at
\begin{eqnarray}
\Lambda_{\text{LP}} & = & \Lambda_0 \exp \left\{ \frac{8 \pi^2}{b_4 g_4^2
	(t'_0)} \left[ 1 - \exp \left( \frac{- \pi b_4}{c_2  \hat{d}} \left[ 1 -
\frac{2}{\pi} \tan^{- 1} \left( \frac{F (t'_0)}{\hat{d}} - \frac{c_1}{2 c_2
	\hat{d}} \right) \right] \right) \right] \right\} \notag\\
& = & \Lambda_0 \exp \left\{ \frac{8 \pi^2}{b_4 g_4^2 (t'_0)} \left[ 1 -
\exp \left( \frac{- \pi b_4}{c_2  \hat{d}} \left[ 1 - \frac{2}{\pi} \tan^{-
	1} \left( \frac{\lambda'_2 (t'_0) + \eta \lambda_3 (t'_0)}{\hat{d} g_4^2
	(t'_0)} - \frac{c_1}{2 c_2 \hat{d}} \right) \right] \right) \right] \right\}
\nonumber\\
& = & \Lambda_{\text{LP}}^{(g_4)} \exp \left[ \frac{- 8 \pi^2}{b_4 g_4^2
	(t'_0)} \exp \left( \frac{- \pi b_4}{c_2  \hat{d}} \left[ 1 - \frac{2}{\pi}
\tan^{- 1} \left( \frac{\lambda'_2 (t'_0) + \eta \lambda_3 (t'_0)}{\hat{d}
	g_4^2 (t'_0)} - \frac{c_1}{2 c_2 \hat{d}} \right) \right] \right) \right],\label{eq:LP_analy} 
\end{eqnarray}
with $\Lambda_{\text{LP}}^{(g_4)} = \exp \left( \frac{- 8 \pi^2}{b_4 g_4^2
	(t'_0)} \right)$.

In this approximate solution, we neglect the $\lambda_4$ term and thus the Landau pole is independent with $\lambda_4$.
However, when $\lambda_4$ too large to be neglected, it should be taken into account. Another simplification in our analytical method is to neglect the $c'f_1^2$ term in Eq.~\eqref{eq:dF}. Since $f_1^2$ may become large so that can not be neglected in higher energy scale, the evolutions of $F$ and $f_1$ couple with each other leading a complicated associated differential equations. If we want to get a reliable quantitative result, we should solve the equations numerically.

\bibliographystyle{utphys}
\bibliography{ref}

\providecommand{\href}[2]{#2}\begingroup\raggedright\begin{thebibliography}{10}

\bibitem{Planck:2018vyg}
{\bfseries Planck} Collaboration, N.~Aghanim {\em et~al.}, ``{Planck 2018
  results. VI. Cosmological parameters},''
  \href{http://dx.doi.org/10.1051/0004-6361/201833910}{{\em Astron. Astrophys.}
  {\bfseries 641} (2020) A6}, \href{http://arxiv.org/abs/1807.06209}{{\ttfamily
  arXiv:1807.06209 [astro-ph.CO]}}. [Erratum: Astron.Astrophys. 652, C4
  (2021)].

\bibitem{LUX:2016ggv}
{\bfseries LUX} Collaboration, D.~S. Akerib {\em et~al.}, ``{Results from a
  search for dark matter in the complete LUX exposure},''
  \href{http://dx.doi.org/10.1103/PhysRevLett.118.021303}{{\em Phys. Rev.
  Lett.} {\bfseries 118} (2017) 021303},
  \href{http://arxiv.org/abs/1608.07648}{{\ttfamily arXiv:1608.07648
  [astro-ph.CO]}}.

\bibitem{XENON:2018voc}
{\bfseries XENON} Collaboration, E.~Aprile {\em et~al.}, ``{Dark Matter Search
  Results from a One Ton-Year Exposure of XENON1T},''
  \href{http://dx.doi.org/10.1103/PhysRevLett.121.111302}{{\em Phys. Rev.
  Lett.} {\bfseries 121} (2018) 111302},
  \href{http://arxiv.org/abs/1805.12562}{{\ttfamily arXiv:1805.12562
  [astro-ph.CO]}}.

\bibitem{PandaX-4T:2021bab}
{\bfseries PandaX-4T} Collaboration, Y.~Meng {\em et~al.}, ``{Dark Matter
  Search Results from the PandaX-4T Commissioning Run},''
  \href{http://dx.doi.org/10.1103/PhysRevLett.127.261802}{{\em Phys. Rev.
  Lett.} {\bfseries 127} (2021) 261802},
  \href{http://arxiv.org/abs/2107.13438}{{\ttfamily arXiv:2107.13438
  [hep-ex]}}.

\bibitem{LUX-ZEPLIN:2022qhg}
{\bfseries LUX-ZEPLIN} Collaboration, J.~Aalbers {\em et~al.}, ``{First Dark
  Matter Search Results from the LUX-ZEPLIN (LZ) Experiment},''
  \href{http://arxiv.org/abs/2207.03764}{{\ttfamily arXiv:2207.03764
  [hep-ex]}}.

\bibitem{Lebedev:2011iq}
O.~Lebedev, H.~M. Lee, and Y.~Mambrini, ``{Vector Higgs-portal dark matter and
  the invisible Higgs},''
  \href{http://dx.doi.org/10.1016/j.physletb.2012.01.029}{{\em Phys. Lett. B}
  {\bfseries 707} (2012) 570--576},
  \href{http://arxiv.org/abs/1111.4482}{{\ttfamily arXiv:1111.4482 [hep-ph]}}.

\bibitem{Abe:2012hb}
T.~Abe, M.~Kakizaki, S.~Matsumoto, and O.~Seto, ``{Vector WIMP Miracle},''
  \href{http://dx.doi.org/10.1016/j.physletb.2012.05.051}{{\em Phys. Lett. B}
  {\bfseries 713} (2012) 211--215},
  \href{http://arxiv.org/abs/1202.5902}{{\ttfamily arXiv:1202.5902 [hep-ph]}}.

\bibitem{Farzan:2012hh}
Y.~Farzan and A.~R. Akbarieh, ``{VDM: A model for Vector Dark Matter},''
  \href{http://dx.doi.org/10.1088/1475-7516/2012/10/026}{{\em JCAP} {\bfseries
  10} (2012) 026}, \href{http://arxiv.org/abs/1207.4272}{{\ttfamily
  arXiv:1207.4272 [hep-ph]}}.

\bibitem{Baek:2012se}
S.~Baek, P.~Ko, W.-I. Park, and E.~Senaha, ``{Higgs Portal Vector Dark Matter :
  Revisited},'' \href{http://dx.doi.org/10.1007/JHEP05(2013)036}{{\em JHEP}
  {\bfseries 05} (2013) 036}, \href{http://arxiv.org/abs/1212.2131}{{\ttfamily
  arXiv:1212.2131 [hep-ph]}}.

\bibitem{Domingo:2013tna}
F.~Domingo, O.~Lebedev, Y.~Mambrini, J.~Quevillon, and A.~Ringwald, ``{More on
  the Hypercharge Portal into the Dark Sector},''
  \href{http://dx.doi.org/10.1007/JHEP09(2013)020}{{\em JHEP} {\bfseries 09}
  (2013) 020}, \href{http://arxiv.org/abs/1305.6815}{{\ttfamily arXiv:1305.6815
  [hep-ph]}}.

\bibitem{Yu:2014pra}
J.-H. Yu, ``{Vector Fermion-Portal Dark Matter: Direct Detection and Galactic
  Center Gamma-Ray Excess},''
  \href{http://dx.doi.org/10.1103/PhysRevD.90.095010}{{\em Phys. Rev. D}
  {\bfseries 90} (2014) 095010},
  \href{http://arxiv.org/abs/1409.3227}{{\ttfamily arXiv:1409.3227 [hep-ph]}}.

\bibitem{Chen:2014cbt}
C.-R. Chen, Y.-K. Chu, and H.-C. Tsai, ``{An Elusive Vector Dark Matter},''
  \href{http://dx.doi.org/10.1016/j.physletb.2014.12.043}{{\em Phys. Lett. B}
  {\bfseries 741} (2015) 205--209},
  \href{http://arxiv.org/abs/1410.0918}{{\ttfamily arXiv:1410.0918 [hep-ph]}}.

\bibitem{Duch:2015jta}
M.~Duch, B.~Grzadkowski, and M.~McGarrie, ``{A stable Higgs portal with vector
  dark matter},'' \href{http://dx.doi.org/10.1007/JHEP09(2015)162}{{\em JHEP}
  {\bfseries 09} (2015) 162}, \href{http://arxiv.org/abs/1506.08805}{{\ttfamily
  arXiv:1506.08805 [hep-ph]}}.

\bibitem{DiFranzo:2015nli}
A.~DiFranzo, P.~J. Fox, and T.~M.~P. Tait, ``{Vector Dark Matter through a
  Radiative Higgs Portal},''
  \href{http://dx.doi.org/10.1007/JHEP04(2016)135}{{\em JHEP} {\bfseries 04}
  (2016) 135}, \href{http://arxiv.org/abs/1512.06853}{{\ttfamily
  arXiv:1512.06853 [hep-ph]}}.

\bibitem{Azevedo:2018oxv}
D.~Azevedo, M.~Duch, B.~Grzadkowski, D.~Huang, M.~Iglicki, and R.~Santos,
  ``{Testing scalar versus vector dark matter},''
  \href{http://dx.doi.org/10.1103/PhysRevD.99.015017}{{\em Phys. Rev. D}
  {\bfseries 99} (2019) 015017},
  \href{http://arxiv.org/abs/1808.01598}{{\ttfamily arXiv:1808.01598
  [hep-ph]}}.

\bibitem{Mohamadnejad:2019vzg}
A.~Mohamadnejad, ``{Gravitational waves from scale-invariant vector dark matter
  model: Probing below the neutrino-floor},''
  \href{http://dx.doi.org/10.1140/epjc/s10052-020-7756-6}{{\em Eur. Phys. J. C}
  {\bfseries 80} (2020) 197}, \href{http://arxiv.org/abs/1907.08899}{{\ttfamily
  arXiv:1907.08899 [hep-ph]}}.

\bibitem{Glaus:2019itb}
S.~Glaus, M.~M\"uhlleitner, J.~M\"uller, S.~Patel, and R.~Santos,
  ``{Electroweak Corrections to Dark Matter Direct Detection in a Vector Dark
  Matter Model},'' \href{http://dx.doi.org/10.1007/JHEP10(2019)152}{{\em JHEP}
  {\bfseries 10} (2019) 152}, \href{http://arxiv.org/abs/1908.09249}{{\ttfamily
  arXiv:1908.09249 [hep-ph]}}.

\bibitem{Arcadi:2020jqf}
G.~Arcadi, A.~Djouadi, and M.~Kado, ``{The Higgs-portal for vector dark matter
  and the effective field theory approach: A reappraisal},''
  \href{http://dx.doi.org/10.1016/j.physletb.2020.135427}{{\em Phys. Lett. B}
  {\bfseries 805} (2020) 135427},
  \href{http://arxiv.org/abs/2001.10750}{{\ttfamily arXiv:2001.10750
  [hep-ph]}}.

\bibitem{Delaunay:2020vdb}
C.~Delaunay, T.~Ma, and Y.~Soreq, ``{Stealth decaying spin-1 dark matter},''
  \href{http://dx.doi.org/10.1007/JHEP02(2021)010}{{\em JHEP} {\bfseries 02}
  (2021) 010}, \href{http://arxiv.org/abs/2009.03060}{{\ttfamily
  arXiv:2009.03060 [hep-ph]}}.

\bibitem{Salehian:2020asa}
B.~Salehian, M.~A. Gorji, H.~Firouzjahi, and S.~Mukohyama, ``{Vector dark
  matter production from inflation with symmetry breaking},''
  \href{http://dx.doi.org/10.1103/PhysRevD.103.063526}{{\em Phys. Rev. D}
  {\bfseries 103} (2021) 063526},
  \href{http://arxiv.org/abs/2010.04491}{{\ttfamily arXiv:2010.04491
  [hep-ph]}}.

\bibitem{Hambye:2008bq}
T.~Hambye, ``{Hidden vector dark matter},''
  \href{http://dx.doi.org/10.1088/1126-6708/2009/01/028}{{\em JHEP} {\bfseries
  01} (2009) 028}, \href{http://arxiv.org/abs/0811.0172}{{\ttfamily
  arXiv:0811.0172 [hep-ph]}}.

\bibitem{Boehm:2014bia}
C.~Boehm, M.~J. Dolan, and C.~McCabe, ``{A weighty interpretation of the
  Galactic Centre excess},''
  \href{http://dx.doi.org/10.1103/PhysRevD.90.023531}{{\em Phys. Rev. D}
  {\bfseries 90} (2014) 023531},
  \href{http://arxiv.org/abs/1404.4977}{{\ttfamily arXiv:1404.4977 [hep-ph]}}.

\bibitem{Gross:2015cwa}
C.~Gross, O.~Lebedev, and Y.~Mambrini, ``{Non-Abelian gauge fields as dark
  matter},'' \href{http://dx.doi.org/10.1007/JHEP08(2015)158}{{\em JHEP}
  {\bfseries 08} (2015) 158}, \href{http://arxiv.org/abs/1505.07480}{{\ttfamily
  arXiv:1505.07480 [hep-ph]}}.

\bibitem{Karam:2015jta}
A.~Karam and K.~Tamvakis, ``{Dark matter and neutrino masses from a
  scale-invariant multi-Higgs portal},''
  \href{http://dx.doi.org/10.1103/PhysRevD.92.075010}{{\em Phys. Rev. D}
  {\bfseries 92} (2015) 075010},
  \href{http://arxiv.org/abs/1508.03031}{{\ttfamily arXiv:1508.03031
  [hep-ph]}}.

\bibitem{Khoze:2016zfi}
V.~V. Khoze and A.~D. Plascencia, ``{Dark Matter and Leptogenesis Linked by
  Classical Scale Invariance},''
  \href{http://dx.doi.org/10.1007/JHEP11(2016)025}{{\em JHEP} {\bfseries 11}
  (2016) 025}, \href{http://arxiv.org/abs/1605.06834}{{\ttfamily
  arXiv:1605.06834 [hep-ph]}}.

\bibitem{Baouche:2021wwa}
N.~Baouche, A.~Ahriche, G.~Faisel, and S.~Nasri, ``{Phenomenology of the hidden
  SU(2) vector dark matter model},''
  \href{http://dx.doi.org/10.1103/PhysRevD.104.075022}{{\em Phys. Rev. D}
  {\bfseries 104} (2021) 075022},
  \href{http://arxiv.org/abs/2105.14387}{{\ttfamily arXiv:2105.14387
  [hep-ph]}}.

\bibitem{Borah:2021ftr}
D.~Borah, A.~Dasgupta, and S.~K. Kang, ``{A first order dark SU(2)$_{D}$ phase
  transition with vector dark matter in the light of NANOGrav 12.5 yr data},''
  \href{http://dx.doi.org/10.1088/1475-7516/2021/12/039}{{\em JCAP} {\bfseries
  12} (2021) 039}, \href{http://arxiv.org/abs/2109.11558}{{\ttfamily
  arXiv:2109.11558 [hep-ph]}}.

\bibitem{Baek:2013dwa}
S.~Baek, P.~Ko, and W.-I. Park, ``{Hidden sector monopole, vector dark matter
  and dark radiation with Higgs portal},''
  \href{http://dx.doi.org/10.1088/1475-7516/2014/10/067}{{\em JCAP} {\bfseries
  10} (2014) 067}, \href{http://arxiv.org/abs/1311.1035}{{\ttfamily
  arXiv:1311.1035 [hep-ph]}}.

\bibitem{Khoze:2014woa}
V.~V. Khoze and G.~Ro, ``{Dark matter monopoles, vectors and photons},''
  \href{http://dx.doi.org/10.1007/JHEP10(2014)061}{{\em JHEP} {\bfseries 10}
  (2014) 061}, \href{http://arxiv.org/abs/1406.2291}{{\ttfamily arXiv:1406.2291
  [hep-ph]}}.

\bibitem{Ghosh:2020ipy}
T.~Ghosh, H.-K. Guo, T.~Han, and H.~Liu, ``{Electroweak phase transition with
  an SU(2) dark sector},''
  \href{http://dx.doi.org/10.1007/JHEP07(2021)045}{{\em JHEP} {\bfseries 07}
  (2021) 045}, \href{http://arxiv.org/abs/2012.09758}{{\ttfamily
  arXiv:2012.09758 [hep-ph]}}.

\bibitem{Nomura:2020zlm}
T.~Nomura, H.~Okada, and S.~Yun, ``{Vector dark matter from a gauged SU(2)
  symmetry},'' \href{http://dx.doi.org/10.1007/JHEP06(2021)122}{{\em JHEP}
  {\bfseries 06} (2021) 122}, \href{http://arxiv.org/abs/2012.11377}{{\ttfamily
  arXiv:2012.11377 [hep-ph]}}.

\bibitem{Hu:2021pln}
Z.~Hu, C.~Cai, Y.-L. Tang, Z.-H. Yu, and H.-H. Zhang, ``{Vector dark matter
  from split SU(2) gauge bosons},''
  \href{http://dx.doi.org/10.1007/JHEP07(2021)089}{{\em JHEP} {\bfseries 07}
  (2021) 089}, \href{http://arxiv.org/abs/2103.00220}{{\ttfamily
  arXiv:2103.00220 [hep-ph]}}.

\bibitem{Cai:2021wmu}
C.~Cai and H.-H. Zhang, ``{Vector dark matter production from catalyzed
  annihilation},'' \href{http://dx.doi.org/10.1007/JHEP01(2022)099}{{\em JHEP}
  {\bfseries 01} (2022) 099}, \href{http://arxiv.org/abs/2107.13475}{{\ttfamily
  arXiv:2107.13475 [hep-ph]}}.

\bibitem{Belyaev:2022shr}
A.~Belyaev, A.~Deandrea, S.~Moretti, L.~Panizzi, and N.~Thongyoi, ``{A
  Fermionic Portal to Vector Dark Matter from a New Gauge Sector},''
  \href{http://arxiv.org/abs/2204.03510}{{\ttfamily arXiv:2204.03510
  [hep-ph]}}.

\bibitem{Elahi:2022hgj}
F.~Elahi and S.~Khatibi, ``{Light Non-Abelian Vector Dark Matter Produced
  Through Vector Misalignment},''
  \href{http://arxiv.org/abs/2204.04012}{{\ttfamily arXiv:2204.04012
  [hep-ph]}}.

\bibitem{Barman:2018esi}
B.~Barman, S.~Bhattacharya, and M.~Zakeri, ``{Multipartite Dark Matter in
  $SU(2)_N$ extension of Standard Model and signatures at the LHC},''
  \href{http://dx.doi.org/10.1088/1475-7516/2018/09/023}{{\em JCAP} {\bfseries
  09} (2018) 023}, \href{http://arxiv.org/abs/1806.01129}{{\ttfamily
  arXiv:1806.01129 [hep-ph]}}.

\bibitem{Barman:2019lvm}
B.~Barman, S.~Bhattacharya, and M.~Zakeri, ``{Non-Abelian Vector Boson as FIMP
  Dark Matter},'' \href{http://dx.doi.org/10.1088/1475-7516/2020/02/029}{{\em
  JCAP} {\bfseries 02} (2020) 029},
  \href{http://arxiv.org/abs/1905.07236}{{\ttfamily arXiv:1905.07236
  [hep-ph]}}.

\bibitem{Chowdhury:2021tnm}
T.~A. Chowdhury and S.~Saad, ``{Non-Abelian vector dark matter and lepton
  g-2},'' \href{http://dx.doi.org/10.1088/1475-7516/2021/10/014}{{\em JCAP}
  {\bfseries 10} (2021) 014}, \href{http://arxiv.org/abs/2107.11863}{{\ttfamily
  arXiv:2107.11863 [hep-ph]}}.

\bibitem{Cirelli:2005uq}
M.~Cirelli, N.~Fornengo, and A.~Strumia, ``{Minimal dark matter},''
  \href{http://dx.doi.org/10.1016/j.nuclphysb.2006.07.012}{{\em Nucl. Phys. B}
  {\bfseries 753} (2006) 178--194},
  \href{http://arxiv.org/abs/hep-ph/0512090}{{\ttfamily arXiv:hep-ph/0512090}}.

\bibitem{Cai:2015kpa}
C.~Cai, Z.-M. Huang, Z.~Kang, Z.-H. Yu, and H.-H. Zhang, ``{Perturbativity
  Limits for Scalar Minimal Dark Matter with Yukawa Interactions: Septuplet},''
  \href{http://dx.doi.org/10.1103/PhysRevD.92.115004}{{\em Phys. Rev. D}
  {\bfseries 92} (2015) 115004},
  \href{http://arxiv.org/abs/1510.01559}{{\ttfamily arXiv:1510.01559
  [hep-ph]}}.

\bibitem{Hamada:2015bra}
Y.~Hamada, K.~Kawana, and K.~Tsumura, ``{Landau pole in the Standard Model with
  weakly interacting scalar fields},''
  \href{http://dx.doi.org/10.1016/j.physletb.2015.05.072}{{\em Phys. Lett. B}
  {\bfseries 747} (2015) 238--244},
  \href{http://arxiv.org/abs/1505.01721}{{\ttfamily arXiv:1505.01721
  [hep-ph]}}.

\bibitem{LZ:2015kxe}
{\bfseries LZ} Collaboration, D.~S. Akerib {\em et~al.}, ``{LUX-ZEPLIN (LZ)
  Conceptual Design Report},''
  \href{http://arxiv.org/abs/1509.02910}{{\ttfamily arXiv:1509.02910
  [physics.ins-det]}}.

\bibitem{MAGIC:2016xys}
{\bfseries MAGIC, Fermi-LAT} Collaboration, M.~L. Ahnen {\em et~al.}, ``{Limits
  to Dark Matter Annihilation Cross-Section from a Combined Analysis of MAGIC
  and Fermi-LAT Observations of Dwarf Satellite Galaxies},''
  \href{http://dx.doi.org/10.1088/1475-7516/2016/02/039}{{\em JCAP} {\bfseries
  02} (2016) 039}, \href{http://arxiv.org/abs/1601.06590}{{\ttfamily
  arXiv:1601.06590 [astro-ph.HE]}}.

\bibitem{CTAConsortium:2012fwj}
{\bfseries CTA Consortium} Collaboration, M.~Doro {\em et~al.}, ``{Dark Matter
  and Fundamental Physics with the Cherenkov Telescope Array},''
  \href{http://dx.doi.org/10.1016/j.astropartphys.2012.08.002}{{\em Astropart.
  Phys.} {\bfseries 43} (2013) 189--214},
  \href{http://arxiv.org/abs/1208.5356}{{\ttfamily arXiv:1208.5356
  [astro-ph.IM]}}.

\bibitem{Elias-Miro:2012eoi}
J.~Elias-Miro, J.~R. Espinosa, G.~F. Giudice, H.~M. Lee, and A.~Strumia,
  ``{Stabilization of the Electroweak Vacuum by a Scalar Threshold Effect},''
  \href{http://dx.doi.org/10.1007/JHEP06(2012)031}{{\em JHEP} {\bfseries 06}
  (2012) 031}, \href{http://arxiv.org/abs/1203.0237}{{\ttfamily arXiv:1203.0237
  [hep-ph]}}.

\bibitem{Kannike:2012pe}
K.~Kannike, ``{Vacuum Stability Conditions From Copositivity Criteria},''
  \href{http://dx.doi.org/10.1140/epjc/s10052-012-2093-z}{{\em Eur. Phys. J. C}
  {\bfseries 72} (2012) 2093}, \href{http://arxiv.org/abs/1205.3781}{{\ttfamily
  arXiv:1205.3781 [hep-ph]}}.

\bibitem{ParticleDataGroup:2020ssz}
{\bfseries Particle Data Group} Collaboration, P.~A. Zyla {\em et~al.},
  ``{Review of Particle Physics},''
  \href{http://dx.doi.org/10.1093/ptep/ptaa104}{{\em PTEP} {\bfseries 2020}
  (2020) 083C01}.

\bibitem{Hambye:1996wb}
T.~Hambye and K.~Riesselmann, ``{Matching conditions and Higgs mass upper
  bounds revisited},'' \href{http://dx.doi.org/10.1103/PhysRevD.55.7255}{{\em
  Phys. Rev. D} {\bfseries 55} (1997) 7255--7262},
  \href{http://arxiv.org/abs/hep-ph/9610272}{{\ttfamily arXiv:hep-ph/9610272}}.

\bibitem{Alloul:2013bka}
A.~Alloul, N.~D. Christensen, C.~Degrande, C.~Duhr, and B.~Fuks, ``{FeynRules
  2.0 - A complete toolbox for tree-level phenomenology},''
  \href{http://dx.doi.org/10.1016/j.cpc.2014.04.012}{{\em Comput. Phys.
  Commun.} {\bfseries 185} (2014) 2250--2300},
  \href{http://arxiv.org/abs/1310.1921}{{\ttfamily arXiv:1310.1921 [hep-ph]}}.

\bibitem{Belanger:2014vza}
G.~B\'elanger, F.~Boudjema, A.~Pukhov, and A.~Semenov, ``{micrOMEGAs4.1: two
  dark matter candidates},''
  \href{http://dx.doi.org/10.1016/j.cpc.2015.03.003}{{\em Comput. Phys.
  Commun.} {\bfseries 192} (2015) 322--329},
  \href{http://arxiv.org/abs/1407.6129}{{\ttfamily arXiv:1407.6129 [hep-ph]}}.

\bibitem{ATLAS:2019nkf}
{\bfseries ATLAS} Collaboration, G.~Aad {\em et~al.}, ``{Combined measurements
  of Higgs boson production and decay using up to $80$ fb$^{-1}$ of
  proton-proton collision data at $\sqrt{s}=$ 13 TeV collected with the ATLAS
  experiment},'' \href{http://dx.doi.org/10.1103/PhysRevD.101.012002}{{\em
  Phys. Rev. D} {\bfseries 101} (2020) 012002},
  \href{http://arxiv.org/abs/1909.02845}{{\ttfamily arXiv:1909.02845
  [hep-ex]}}.

\bibitem{CMS:2018uag}
{\bfseries CMS} Collaboration, A.~M. Sirunyan {\em et~al.}, ``{Combined
  measurements of Higgs boson couplings in proton\textendash{}proton collisions
  at $\sqrt{s}=13\,\text {Te}\text {V} $},''
  \href{http://dx.doi.org/10.1140/epjc/s10052-019-6909-y}{{\em Eur. Phys. J. C}
  {\bfseries 79} (2019) 421}, \href{http://arxiv.org/abs/1809.10733}{{\ttfamily
  arXiv:1809.10733 [hep-ex]}}.

\end{thebibliography}\endgroup

\end{document}